\documentclass[reqno, 11pt]{amsart}
\usepackage[bottom]{footmisc}
\usepackage[percent]{overpic}
\usepackage{calc,graphicx,amsfonts,amsthm,amscd,amsmath,amssymb,enumerate,dsfont,mathrsfs,
paralist,bbm,tikz,pdfsync}
\usepackage{subfig}

\usetikzlibrary{patterns,arrows}
\usetikzlibrary{shapes,arrows,spy,positioning,snakes}

\usetikzlibrary{lindenmayersystems}
\usetikzlibrary{calc}
\usepackage{stmaryrd}
\usepackage[numeric,initials,nobysame]{amsrefs}
\usepackage{hyperref}

\usepackage{booktabs}
\newcommand{\ie}{\hbox{\it i.e.\ }}

\reversemarginpar
\newlength\fullwidth
\numberwithin{equation}{section}

\DeclareMathSymbol{\leqslant}{\mathalpha}{AMSa}{"36} 
\DeclareMathSymbol{\geqslant}{\mathalpha}{AMSa}{"3E} 
\DeclareMathSymbol{\eset}{\mathalpha}{AMSb}{"3F}     
\renewcommand{\leq}{\;\leqslant\;}                   
\renewcommand{\geq}{\;\geqslant\;}                   
\newcommand{\suptwo}[2]{\sup_{\substack{#1 \\ #2}}} 
\newcommand{\sumtwo}[2]{\sum_{\substack{#1 \\ #2}}} 
\renewcommand{\b}{\beta}

\def\1{\ifmmode {1\hskip -3pt \rm{I}} \else {\hbox {$1\hskip -3pt \rm{I}$}}\fi}

\newcommand{\cov}{\operatorname{Cov}}

\newcommand{\D}{\Delta}

\renewcommand{\b}{\beta}
\renewcommand{\l}{\lambda}
\renewcommand{\L}{\Lambda}

\renewcommand{\l}{\lambda}
\renewcommand{\a}{\alpha}
\renewcommand{\d}{\delta}
\renewcommand{\t}{\tau}

\newcommand{\g}{\gamma}
\newcommand{\G}{\Gamma}

\newcommand{\e}{\varepsilon}

\newtheorem{theorem}{Theorem}[section]
\newtheorem{lemma}[theorem]{Lemma}
\newtheorem{proposition}[theorem]{Proposition}
\newtheorem{corollary}[theorem]{Corollary}
\newtheorem{remark}[theorem]{Remark}

\newtheorem{claim}[theorem]{Claim}
\newtheorem{definition}[theorem]{Definition}
\newtheorem{maintheorem}{Theorem}

\newtheorem*{question*}{Question}

\newtheorem*{remark*}{Remark}
\newtheorem*{idefinition*}{Definition}
\newtheorem*{example*}{Example}

\newcommand{\cA}{\ensuremath{\mathcal A}}
\newcommand{\cB}{\ensuremath{\mathcal B}}

\newcommand{\cF}{\ensuremath{\mathcal F}}
\newcommand{\cG}{\ensuremath{\mathcal G}}
\newcommand{\cH}{\ensuremath{\mathcal H}}

\newcommand{\cK}{\ensuremath{\mathcal K}}
\newcommand{\cL}{\ensuremath{\mathcal L}}

\newcommand{\cN}{\ensuremath{\mathcal N}}
\newcommand{\cO}{\ensuremath{\mathcal O}}
\newcommand{\cP}{\ensuremath{\mathcal P}}

\newcommand{\cR}{\ensuremath{\mathcal R}}
\newcommand{\cS}{\ensuremath{\mathcal S}}
\newcommand{\cT}{\ensuremath{\mathcal T}}

\newcommand{\cZ}{\ensuremath{\mathcal Z}}

\newcommand{\bbE}{{\ensuremath{\mathbb E}} }
\newcommand{\bbF}{{\ensuremath{\mathbb F}} }

\newcommand{\bbL}{{\ensuremath{\mathbb L}} }

\newcommand{\bbN}{{\ensuremath{\mathbb N}} }

\newcommand{\bbP}{{\ensuremath{\mathbb P}} }

\newcommand{\bbR}{{\ensuremath{\mathbb R}} }

\newcommand{\bbZ}{{\ensuremath{\mathbb Z}} }

\let\a=\alpha \let\b=\beta   \let\d=\delta  \let\e=\varepsilon
 \let\g=\gamma       \let\l=\lambda
          \let\p=\pi  
  \let\s=\sigma \let\t=\tau   
\let\y=\upsilon \let\x=\xi 
\let\D=\Delta   \let\G=\Gamma  \let\L=\Lambda 
\let\O=\Omega      

\newcommand\T[1]{
  \coordinate (D) at (#1);
  \path let \p1 = (D) in coordinate (t) at (\x1-0.5*\y1,{0.5*sqrt(3)*\y1});
}

\def\dtri#1{
\T{#1}
\begin{scope}[shift=(t)]
  \draw [fill=white,opacity=0.5](0,0) -- (0.5,{0.5*sqrt(3)}) -- (-0.5,{0.5*sqrt(3)}) -- cycle;
\end{scope}
}

\def\Btri#1{
\T{#1}
\begin{scope}[shift=(t)]
  \draw [fill=gray](0,0) -- (0.5,{0.5*sqrt(3)}) -- (-0.5,{0.5*sqrt(3)}) -- cycle;
\end{scope}
}

\def\Bltri#1{
\T{#1}
\begin{scope}[shift=(t)]
  \draw [fill=lightgray](0,0) -- (0.5,{0.5*sqrt(3)}) -- (-0.5,{0.5*sqrt(3)}) -- cycle;
\end{scope}
}
\def\Bsq#1{
\begin{scope}[shift=(#1)]
  \draw [fill=gray](0,0) -- (1,1) -- (0,1) -- cycle;
\end{scope}
}

\def\Blsq#1{
\begin{scope}[shift=(#1)]
  \draw [fill=lightgray](0,0) -- (1,1) -- (0,1) -- cycle;
\end{scope}
}

\newcommand{\rosso}{\textcolor{black}}

\begin{document}
\title{Mixing length scales of low temperature spin plaquettes models } 
\author[P. Chleboun]{P. Chleboun}
 \address{P. Chleboun, Department of Statistics, University of Oxford, 24-29 St Giles', Oxford, OX1 3LB, United Kingdom} 
\email{paul.chleboun@stats.ox.ac.uk}
\author[A. Faggionato]{A. Faggionato}
\address{A. Faggionato, Dipartimento di  Matematica, Universit\`a  La  Sapienza, P.le Aldo Moro  2, 00185  Roma, Italy. }\email{faggiona@mat.uniroma1.it}
\author[F. Martinelli]{F. Martinelli}
\address{F. Martinelli, Dipartimento di Matematica e Fisica, Universit\`a Roma Tre, Largo S.L. Murialdo 00146, Roma, Italy}
\email{martin@mat.uniroma3.it}
\author[C. Toninelli]{C. Toninelli}
\address{C. Toninelli, Laboratoire de Probabilit\'es et Mod\`eles Al\`eatoires
  CNRS-UMR 7599 Universit\'es Paris VI-VII 4, Place Jussieu F-75252 Paris Cedex 05 France}
\email{cristina.toninelli@upmc.fr}
\thanks{We acknowledge the support by the ERC Starting Grant 680275
  MALIG and the PRIN 20155PAWZB "Large Scale Random Structures"}
\begin{abstract}
Plaquette models are short range ferromagnetic spin models that play a key role in the dynamic facilitation approach to the liquid glass transition. 
In this paper we perform  a rigorous study of the thermodynamic properties of two dimensional plaquette models,
the square and triangular plaquette models.
 We prove that for any positive temperature both models have a unique infinite volume Gibbs measure with exponentially decaying correlations.  
We analyse the scaling of three a priori different static correlation lengths in the small temperature regime, the mixing, cavity and multispin correlation lengths.
Finally, using the symmetries of the model we determine an exact self
similarity property for the infinite volume Gibbs measure.
\end{abstract}
\maketitle
\section{Introduction}
Providing a clear and deep understanding of the liquid glass transition and of the glassy state of matter remains an open challenge for condensed matter physicists (see \cite{BiroliBerthier} for a review on the various theoretical approaches).
One of the theories of the glass transition,
known as {\sl dynamical facilitation} (DF),  relies on two basic assumptions.
 The first paradigm is that the dominant relaxation mechanism when approaching the glass transition
should be facilitated relaxation: a local region that relaxes allows (facilitates) a neighbouring region to relax as well. The second assumption is that slowing down should be due to a decreasing density of these local facilitation regions: mobility is sparse in a low temperature (dense) liquid. Assuming that 
any local relaxation events needs to be triggered by a nearby relaxation implies that mobility is essentially conserved and  propagates through the system. These are strong assumptions which lack a clearcut experimental validation. Indeed, experimental tests are difficult in the absence of a systematic coarse-graining procedure to represent a molecular model with continuous degrees of freedom in terms of mobility variables. However, an undiscussed success of the DF scenario is that it has lead to the introduction of different models which display several key properties of glassy dynamics. In turn, these models have provided 
a deeper understanding of  the dynamical heterogeneities that occur in supercooled liquids: fast and slow regions coexist and their typical size increases while decreasing temperature.

The first class of models that has been studied in the context of the DF scenario are the so called Kinetically Constrained Models (KCM), which feature  trivial statics and constrained dynamics. KCM  include for example the East and Friedrickson-Andersen model
(see  \cite{GarrahanSollichToninelli}, \cite{EastFMRT}  and \cite{CFM} for some references to the physical and mathematical literature respectively).
Two crucial difficulties in justifying KCM as models for the liquid glass transition are the following: (i) 
it is not clear how kinetic constraints can truly emerge from
the unconstrained dynamics of a many-body system; (ii) 
since KCM have a trivial thermodynamics they cannot account for the eventual growth of a static amorphous order. To cope with both problems, a second class of models has been introduced, the so called {\sl plaquette models} \cite{Newman,Garrahanreview}. 

In this paper we focus on the
 two dimensional plaquette models: the {\sl square plaquette model} (SPM) \cite{Garrahanreview} on $\mathbb Z^2$ and the {\sl triangular plaquette model} (TPM) \cite{Newman} on the triangular lattice $\mathcal T$.
Both SPM and TPM are systems of $\pm 1$ spins with short range non disordered ferromagnetic interactions.  
In order to define the Hamiltonian and therefore the Gibbs measure, we
need first to define the plaquettes and the plaquette variables.
For the SPM the plaquettes are all the unit squares in $\mathbb Z^2$ while
for the TPM they are all the downward-pointing unit
side triangles in $\cT$. For a given spin configuration
$\sigma\in\{\pm 1\}^{\mathbb Z^2}$ or $\sigma\in\{\pm 1\}^{\mathcal
  T}$ and a given plaquette $\cP$, the corresponding {\sl plaquette
  variable} is defined as $\prod_{x\in \cP}\s_x$. Then, the Hamiltonian is defined as
$-1/2$ times the sum of  all plaquette variables. Despite the 
non interacting form of the Hamiltonian in terms of the plaquette variables, 
the thermodynamics  of the spin variables is non trivial. Indeed, the
correspondence among the spin and plaquette variables is not
one-to-one, and the spin ground state is highly degenerate (see
\cite{Garrahanreview} for an informal description of the involved
symmetries). In particular, both models feature diverging static
correlation lengths as the temperature tends to zero \cite{BerthierGarrahanJack,GarrahanJack}.

The study of the dynamics of  SPM and TPM under the natural single spin flip Monte Carlo or Metropolis dynamics has been the focus of several works in physics literature. 
Numerical simulations clearly indicate the occurrence of glassy dynamics at low temperature \cite{BerthierGarrahanJack,Newman,GarrahanJack,Garrahanreview}.
In this regime the dynamics in terms of plaquette variables  is usually described via an  effective dynamics of free defects  subject to kinetic constraints. Indeed, flipping a single spin changes the value of all the plaquette variables containing the corresponding site. 
Thus plaquettes with variable equal to $-1$, the so called defects, are stable when they are isolated.
Furthermore, relaxation is dominated by flips occurring in the vicinity of defects.
This shows that effective kinetic constraints can  naturally emerge from many body interactions. Concerning the critical scaling of  time scales, heuristic analysis and numerical simulations clearly indicate an Arrhenius scaling for SPM  and a super Arrhenius scaling for TPM \cite{BerthierGarrahanJack,Newman}. 
This difference is due to the nature of the energy barriers that should be overcome to bring isolated defects together and annihilate them.

The main focus of our paper is a rigorous study of the thermodynamic properties of SPM and TPM.
Our first result (cf. Theorem \ref{thm:main1}) proves that, for any positive
temperature, both models have a unique infinite volume Gibbs measure
with a strong form of spatial mixing. For SPM the uniqueness of the
Gibbs measure was known \cite{FS}*{Theorem 3.2} but not the strong spatial
mixing. 
Notice that for the SPM transfer matrix techniques allow for the
exact calculation of the free energy at zero external field
\cites{EspriuPrats,Mueller}.
The techniques we employ, partially adapted from \cite{FS}, are robust
and they can also be applied to cases in which it is not feasible to exactly evaluate the free energy. 

We then analyse the scaling of three natural length scales in the small temperature regime.
The largest scale, $\ell_c^{(mix)}$, measures the critical scale  at
which correlations decay in the
bulk and also close to the boundary, uniformly in the boundary condition. 
We prove that  $\ell_c^{(mix)}$
scales as $e^{\beta}$ modulo polynomial corrections in $\beta$
(Theorem  \ref{thm:main1} and Remark \ref{rem3.10}).

The second scale, $\ell_c^{(cavity)}$, measures the minimal distance at which boundary conditions do not significantly
affect the average of local observables located far from the boundary. For SPM we show that $\ell_c^{(cavity)}$ scales as $e^{\beta}$
modulo polynomial corrections in $\beta$ (Theorem
\ref{thm:main2}). 
For TPM we find $e^{\frac{\ln 2}{\ln 3} \beta}\leq
\ell_c^{(cavity)}\leq \beta^2 e^{\beta}$. 
Different types of cavity or
point-to-set correlation lengths have been defined and measured in
plaquette models and other glassy systems.  
These are highly relevant
in connection with the key problem of measuring subtle correlations
due to the growing of an amorphous order in the glass state
(references in \cite{BiroliBerthier}).
\begin{remark}
The scale $\ell_c^{(cavity)}$ does not correspond  to the cavity correlation
length that can be measured via static overlap functions but rather to
the a priori larger length beyond which the cavity behaves as the
bulk. 
In fact, for the SPM case, $\ell_c^{(cavity)}$ scales as the second crossover length in \cite{GarrahanJack}.  
In turn, this length is expected to diverge as the dynamical length that can be extracted from the four point correlator (see also \cite{BerthierGarrahanJack}) which is relevant for the study of dynamical heterogeneities.
For the TPM  it has been conjectured from numerical simulations that there exists a unique static and dynamic correlation length scaling with $e^{\frac{\ln 2}{\ln 3}\beta}$ \cite{GarrahanJack}. 
This would suggest that our lower bound for $\ell_c^{(cavity)}$ is the correct bound. 
However, since we are taking the supremum over all boundary conditions
in the definition of $\ell_c^{(cavity)}$ a larger scaling could occur. 
Notice also that our results  prove that there is another
critical length, $\ell_c^{(mix)}$, with a faster divergence. 
\end{remark}
The third and  smallest scale, $\ell_c^{(multispin)}$,  is the correlation length for the
product of spin variables in the infinite volume Gibbs measure. 
We find (cf. Theorem  \ref{thm:main2}) that $\ell_c^{(multispin)}\sim
e^{\beta/2}$ for the SPM and $\ell^{(multispin)}_c\sim e^{\frac{\ln
    2}{\ln 3} \beta}$ for the TPM.
Both scalings  were previously derived in physical literature (see \cite{BerthierGarrahanJack,Garrahanreview}) for special sets of spins, namely located at the vertices of an equilateral triangle for the TPM and of a rectangle for the SPM. 

Finally, using the symmetries of the model we determine an exact self
similarity property for infinite volume Gibbs measure (cf. Theorem \ref{thm:main3}).

\section{Notation and Models}
\subsection{Notation}
For an integer $n$ we will write $[n]$ for the set
$\{1,\dots,n\}$ and for any finite set $\L\subset \bbZ^d$, we will write
$|\L|$ for its cardinality. The $\ell_1$-distance in $\bbZ^d$ will be
denoted by $d(\cdot,\cdot)$ and the standard basis vectors by $\vec e_1,\dots,\vec e_d$. Sometime we will write $x_1,x_2$
for the coordinates of a vertex $x\in \bbZ^2$. 
Given $\L \subset \bbZ^d$ the basic configuration space will be $\O_\L:=  \{ -1, 1\}^{\L}$. When
$\L=\bbZ^d$ we drop the corresponding suffix from our notation. Given
$\s\in \O$ we let $\s_x$ be its value at the site $x$, in the
sequel the \emph{spin at $x$}, and for any finite $V$, we let
$\s_V =(\s_x)_{x\in V}$ and $[\s]_V:= \prod_{x \in V} \s_x$.  Given
  two disjoint subsets $V,V'$ of $\bbZ^d$, and two configurations
  $\s\in\O_{V}$ and $\s' \in \O_{V'}$ we denote by $\s \otimes \s'$
  the configuration in  $\O_{V\cup V'}$ whose restrictions to $V$ and
    $V'$ coincide with $\s$ and $\s'$ respectively. 
For any $\s\in \O$ and vertex $x\in \bbZ^2$,  $\s^x$ will denote
  the configuration obtained from $\s$ by flipping the spin at
  $x$. Given a function $f:\O\mapsto \bbR$ the smallest set $V$ such that $f$ does not
depend on $\s_{V^c}$ will be called the support of $f$ and it will be
denoted by $S_f$. A function $f$ is said to be \emph{local} if its
support is finite.  

Finally we recall that $f(x) =O(g(x))$ as $x \to +\infty$ means that $|f(x)|\le Cg(x) $ for some
  constant $C$ and any $x$ large enough and that $f=\O(g)$ if $g=O(f)$. We will also
  write $f\asymp g$ if $f=O(g)$ and $g=O(f)$.

\subsection{Finite volume Gibbs measures}
\label{sec:Gibbs measure}In order to define the finite volume Gibbs measure of our models with
spin boundary conditions $\tau\in \O$, it will be convenient to first introduce the
notation $\O_\L^\t$ to denote those configurations $\s\in \O$ such
that $\s_{\L^c}=\t_{\L^c}$, where $\t\in \O$ and $\L\subset \bbZ^d$.  
Let  $B_*$ be a finite subset of $\bbZ^d$, in the sequel referred to as the
\emph{fundamental plaquette}, and let $\cB:= \{ B_*+x \,:\, x \in
\bbZ^d\}$. Elements of the collection $\cB$ will be called
\emph{plaquettes}. Given $\L\subset \bbZ^d$ we also let $\cB(\L)= \{B\in
\cB:\ B\cap \L\neq \emptyset\}$. Finally, given a finite set $\L$ and
$\t\in \O$, we
define the Gibbs measure in $\L$ with boundary conditions $\t$
and fundamental plaquette $B_*$ as the
positive probability measure $\mu_\L^{\b,\t}$ on $\O_\L^\t$ given by
\begin{equation*}
\mu_\L ^{\b,\t} (\s):= \frac{ e^{\frac{\b}{2} \sum_{B\in \cB(\L)}[\s]_B}}{ Z^{\b,\t}_\L } \,,\qquad \s \in \O^\t_\L,
\end{equation*} 
where $\b>0$ is the inverse temperature and the normalisation constant
(partition function) $Z_\L^{\b,\t}$ takes the form 
\begin{equation}\label{africa}
Z_\L^{\b,\t}=\sum_{\s\in \O^\t_\L}e^{\frac{\b}{2}\sum_{B\in \cB(\L)}[\s]_B}\,.
\end{equation}
In the sequel we will denote by
$\cov_\L^{\b,\t}(f,g)$ the covariance w.r.t. $\mu_\L^{\b,\t}$ of two functions
$f,g:\O_\L^\t\mapsto \bbR$. 

Using the fact that the fundamental plaquette $B_*$ is finite,
$\mu_\L^{\b,\t}$ is the finite volume Gibbs measure in $\L$ with boundary
conditions $\t$  for the finite
range, $\pm 1$ spin model on $\bbZ^d$ with (formal) Hamiltonian
$H(\s)=-\frac 12 \sum_{B\in \cB}[\s]_B$. As usual for spin systems we
set $\|H\|= |B_*|/2$. 
\subsubsection{Specific models} 
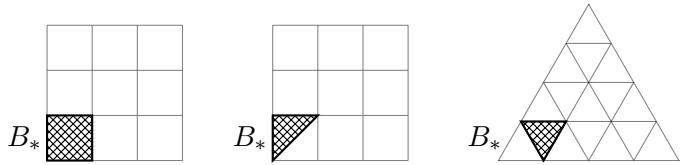
\begin{figure}
\centering
\begin{tikzpicture}[scale=0.6]
\begin{scope}
\draw  [semitransparent] (0,0) grid [step=1] (3,3);
 \draw [thick, pattern=crosshatch]  (0,0) rectangle (1,1);
\node  at (-0.5,0.5) {$B_*$};
\end{scope}
\begin{scope}
\draw  [semitransparent] (5,0) grid [step=1] (8,3);
\draw [thick, pattern=crosshatch] (5,0) -- (6,1) -- (5,1)-- cycle;
\node  at (4.5,0.5) {$B_*$};
\end{scope}
\newcommand*\rows{4}
\begin{scope}[shift={(10,0)}]
    \foreach \row in {0, 1, ...,\rows} {
        \draw [semitransparent] ($\row*(0.5, {0.5*sqrt(3)})$) -- ($(\rows,0)+\row*(-0.5, {0.5*sqrt(3)})$);
        \draw [semitransparent] ($\row*(1, 0)$) -- ($(\rows/2,{\rows/2*sqrt(3)})+\row*(0.5,{-0.5*sqrt(3)})$);
        \draw [semitransparent] ($\row*(1, 0)$) -- ($(0,0)+\row*(0.5,{0.5*sqrt(3)})$);
    }
\draw [thick, pattern=crosshatch] (1,0)--(0.5,{0.5*sqrt(3)})--(1.5,{0.5*sqrt(3)})--cycle;
\node at (-0.3,0.5) {$B_*$};
 \end{scope}
\end{tikzpicture}
\caption{⟨The fundamental plaquette for the SPM (left) and for the TPM in $\bbZ^2$
  (center) and in the triangular lattice $\cT$ (right).⟩} 
\label{fig:1}
\end{figure}

In this paper we will concentrate on
two models in two dimensions: 
the square plaquette model (SPM in the sequel), with fundamental
plaquette $B_*\subset \bbZ^2$ equal to the unit square $\{0,1\}^2$ and the \emph{triangular plaquette
  model} (TPM) with $B_*$ equal to the right triangle with vertices
the origin, $\vec e_2$ and $\vec
e_1+\vec e_2$ (see Figure \ref{fig:1}). 
In
the language of
\cite{FS} both models are \emph{trivial}, \ie each $B\in \cB$ is the
translate of exactly one fundamental plaquette $B_*$. The SPM is also
\emph{factorizable} because $B_*$ is a rectangle, \ie of the form
$B_*= B_*^{(1)}\times B_*^{(2)}$ with $B_*^{(i)}\subset \bbZ$.
\begin{remark}
\label{rem:TPM}
Usually the TPM is defined on the  triangular lattice $\cT$
(cf. e.g. \cite{Newman}) in
which the fundamental plaquette is any downward--pointing unit triangle.
Using the bijection $\Phi$ between $\cT$ and $\bbZ^2$ given by $\cT\ni x= x_1
\vec a_1 +x_2\vec a_2 \mapsto (x_1,x_2)\in \bbZ^2$, where $\vec
a_1=\vec e_1$ and $\vec a_2 = \cos (\frac{2}{3} \pi)\vec e_1 + \sin (
\frac{2}{3} \pi)\vec e_2$, we see immediately that any unit
downward--pointing triangle of $\cT$  is transformed into a plaquette
$B_*+x,\ x \in \bbZ^2$, where $B_*$ is the right triangle above. In
the sequel and if no confusion arises, we will work indifferently
either in $\bbZ^2$ or in $\cT$ according to the geometric convenience.  
\end{remark}
\section{Main results}
\label{sec:main results}
Our results are essentially twofold. The first set of results analyses
the spatial mixing properties of the two models (decay of
correlations, influence of boundary conditions and uniqueness of the
infinite volume Gibbs
measure). The second set of results discusses a self-similarity
property of the
unique infinite volume Gibbs measure.

We first prove that
for all temperatures the SPM and the TPM satisfy the so called
\emph{strong mixing property on regular volumes} (we refer the reader
to \cites{MO1,MO2,MO3} for a
thorough analysis of this property and for 
a critical discussion of its relation with the notion
of complete analyticity of Dobrushin-Shlosman \cites{DS1,DS2}, and to
\cite{M} for a simple and direct approach to prove it). As is
well known the above property implies  the
existence of a unique infinite volume
Gibbs measure $\mu^\b=\lim_{\L\uparrow \bbZ^2}\mu_\L^{\b,\t}$ with exponentially decaying correlations and finite
logarithmic Sobolev constant \cite{MO2}. Already at this level a
first critical scale $\ell_c^{(\text{mix})}$ appears, measuring the minimal distance $\ell$ at which
two functions $f,g$ with supports at distance $\ell$ have a small
covariance in a box $\L$ containing their supports, uniformly in the
boundary conditions $\t$.

Secondly at low temperature we analyse the dependence
on $\b$ of two other natural critical length scales related to the general
concept of spatial mixing. The
first scale, $\ell_c^{(\text{cavity})}$, measures the minimal distance at which boundary conditions
do not significantly affect the average of local observables located far from
the boundary. The second scale, $\ell_c^{(\text{multispin})}$, is the
minimal scale $\ell$ such that the
infinite volume average of multispins separated one from each other by
a distance at least $\ell$ becomes small. 
As we discuss in the appendix these three length scales
are ordered:
\begin{equation}
  \label{eq:ordering}
\ell_c^{(\text{multispin})}=O(\ell_c^{(\text{cavity})})= O(\b\, \ell_c^{(\text{mix})}).
\end{equation}
We will now define precisely the above scales and state the main
results concerning their asymptotics for large $\b$.
\subsection{Spatial Mixing Results}
\label{mixing} 
We begin by recalling the notion of strong mixing on regular
volumes. 
\begin{definition}
Given a positive integer $\ell,$ a finite set $\L\subset \bbZ^2$ is said to be
$\ell$-regular if it is the disjoint union of squares of the form
$Q_\ell+\ell x$, $x\in \bbZ^2$, where $Q_\ell= [\ell]^2$.
 \end{definition}
\begin{definition}
We say that strong mixing on $\ell$-regular sets with positive
constants $C,m$ holds, in the sequel denoted $SM(\ell,C,m)$, if for all
$\ell$-regular sets $\L$, all boundary conditions $\t\in \O$ and all
functions $f,g:\O_\L^\t\mapsto \bbR$ we have
\begin{equation}
  \label{eq:SM1}
|\cov_\L^{\b,\t}(f,g)|\le C \|f\|_\infty \|g\|_\infty |S_f||S_g|e^{-m\, d(S_f,S_g)}.
\end{equation}
\end{definition}
\begin{remark}The key point in the definition of strong mixing is the arbitrariness
of the location of the functions $f,g$. Their supports could in
fact be far from each other but close to
the boundary of $\L$. Thus strong mixing requires the exponential decay of
correlations not only in the bulk but also close to the boundary. It
is known \cite{MOS} that in two dimensions bulk exponential decay
(technically refereed to as \emph{weak mixing}) implies strong mixing
on regular sets,
essentially because the boundary of a square
is one dimensional and information cannot propagate in a one
dimensional space. In higher dimensions there are example of models
\cite{Sh} with bulk exponential decay but no strong mixing property
because of the occurrence of boundary phase transitions.     
\end{remark}
As in the original work of Dobrushin-Shlosman
\cites{DS1,DS2} in order to prove $SM(\ell,C,m)$ it is enough to
verify that a certain \emph{finite volume condition} on boxes
of side $\ell$ holds (see again \cites{MO1,MO3,M}).   
Following e.g. \cite{M}*{Proposition 3.1 and Theorem 3.3}, we have in fact
the following result:
\begin{proposition}[\cite{M}] 
\label{thm:fin-size} Given $x$ let  
\[
h_x(\s)= \exp\bigl(\, \frac{\b}{2}\sumtwo{B\in \cB \,:\,}{B\ni
    x\ }([\s^x]_B-[\s]_B)\bigr).
\] 
Let also 
\[
\varphi(\ell)=\suptwo{x,y\notin Q_\ell}{d(x,y)\ge \ell/4}\sup_\t
|\cov_{Q_\ell}^{\b,\t}(h_x,h_y)|.
\]
Then there exist positive constants $\e_0, c_0>0$ independent of $\b$ such that the inequality
\begin{equation}
  \label{eq:SM}
e^{4\b\|H\|}
\ell
  \,\varphi(\ell)\le \e_0  
\end{equation}
implies $SM(\ell,C,m)$ with $C= e^{c_0
  (\b+1) }$ and $m = 1/\ell$.
\end{proposition}
\begin{remark}
\label{rem:h}The special role played by the functions $\{h_x\}_{x\in \bbZ^2}$ in
checking strong mixing is
due to the following basic identity
\begin{equation}
  \label{eq:h.1}
\mu_\L^{\b,\t^x}(f)-\mu^{\b,\t}_\L(f)=\frac{\cov_\L^{\b,\t}(h_x ,f)}{\mu_\L^{\b,\t}(h_x)}\quad
\forall x\notin \L.
\end{equation}
Thus, if $f$ has a small covariance with $h_x$, its
expectation is not sensitive to a change in the boundary conditions
at $x$. \end{remark}
Notice that in order to verify \eqref{eq:SM} we need some decay
(at least $1/\ell$) of the
function $\varphi(\ell)$ for $\ell$ large enough. The
smallest scale at which the required decay takes place will be our
first critical scale. 
\begin{definition}
We define the \emph{strong mixing scale} $\ell_c^{(\text{mix})}$ as the smallest integer $\ell$ such that
\eqref{eq:SM} holds.    
\end{definition}
Our first theorem then says that strong mixing on regular volumes
holds for both SPM and TPM.
\begin{maintheorem} 
\label{thm:main1}For both SPM and TPM 
  $\ell_c^{(\text{mix})}=O(\b e^\b)$ as $\b\to \infty$.
In particular both models have a unique infinite volume Gibbs measure $\mu^\b$
with exponential decay of correlations and zero magnetisation $\mu^\b(\s_0)=0$. 
\end{maintheorem}
\begin{remark}
It is easily seen that the SPM and TPM do not satisfy
the Dobrushin-Shlosman stronger form of decay of correlations
\cite{DS2} at low temperatures, \ie \eqref{eq:SM1} for
\emph{all} finite sets $\L$ and not just for the $\ell$-regular
ones. Partition $\bbZ^2$ into the odd and even
sub-lattices
and
consider the SPM on the even sub-lattice 
with plus boundary conditions on the odd one. The resulting system is
clearly the standard ferromagnetic Ising model on $\bbZ^2$
for which a phase transition occurs at low temperature. Hence it is not
possible to have \eqref{eq:SM1} at low temperature for all subsets
$\L$ of
the even sub-lattice with plus boundary conditions on the odd one. 

For the TPM the role played
by the odd sub-lattice for the SPM is played by the image under the
mapping $\cT \stackrel{\Phi}\mapsto\bbZ^2$ described in Remark
\ref{rem:TPM} of the subset $\cO=\cT\setminus \cH$, where $\cH$ is the
hexagonal tiling of the plane with each hexagon formed by the union
of six  triangles of $\cT$ (cf. Figure \ref{fig:2}).
\begin{figure}[ht]
\centering  
\begin{tikzpicture}[scale=0.4]
\begin{scope}
\foreach \i/\j in {0/0,1/0,2/0,3/0,0/1,1/1,2/1,3/1,0/2,1/2,2/2,3/2,0/3,1/3,2/3,3/3}{%
\draw [ultra thick] ({0.5*3*\i},{0.5*sqrt(3)*\i+sqrt(3)*\j}) -- ({0.5+0.5*3*\i},{(\i+1)*0.5*sqrt(3)+sqrt(3)*\j}) -- ([turn] -60:1) -- ([turn] -60:1) -- ([turn] -60:1) -- ([turn] -60:1) -- cycle; 
}
\end{scope}
\begin{scope}
\foreach \i/\j in
 {0/0,1/0,2/0,3/0,0/1,1/1,2/1,3/1,0/2,1/2,2/2,3/2,0/3,1/3,2/3,3/3}
  \draw [pattern=dots] ({\i*0.5*3+1},{\i*sqrt(3)*0.5+sqrt(3)*\j}) --
  ({0.5+\i*0.5*3},{sqrt(3)*0.5+\i*sqrt(3)*0.5+sqrt(3)*\j}) --
  ({3*0.5+\i*0.5*3},{sqrt(3)*0.5+\i*sqrt(3)*0.5+sqrt(3)*\j})-- cycle;
\end{scope}
\begin{scope}
\foreach \i/\j in {0/0,1/0,2/0,3/0,0/1,1/1,2/1,3/1,0/2,1/2,2/2,3/2,0/3,1/3,2/3,3/3}
  \draw [pattern=dots] ({\i*0.5*3+0.5},{\i*sqrt(3)*0.5+sqrt(3)*\j-0.5*sqrt(3)}) -- ({\i*0.5*3},{sqrt(3)*0.5+\i*sqrt(3)*0.5+sqrt(3)*\j-0.5*sqrt(3)}) --
  ({3*0.5+\i*0.5*3-0.5},{sqrt(3)*0.5+\i*sqrt(3)*0.5+sqrt(3)*\j-0.5*sqrt(3)})--cycle;
\end{scope}
\begin{scope}
\foreach \i/\j in {0/0,1/0,2/0,3/0,0/1,1/1,2/1,3/1,0/2,1/2,2/2,3/2,0/3,1/3,2/3,3/3}
  \draw [pattern=dots] ({\i*0.5*3+1.5},{\i*sqrt(3)*0.5+sqrt(3)*\j-0.5*sqrt(3)}) -- ({\i*0.5*3+1},{sqrt(3)*0.5+\i*sqrt(3)*0.5+sqrt(3)*\j-0.5*sqrt(3)}) --
  ({3*0.5+\i*0.5*3-0.5+1},{sqrt(3)*0.5+\i*sqrt(3)*0.5+sqrt(3)*\j-0.5*sqrt(3)})--cycle;
\end{scope}
\begin{scope}
\foreach \i/\j in {0/0,1/0,2/0,3/0,0/1,1/1,2/1,3/1,0/2,1/2,2/2,3/2,0/3,1/3,2/3,3/3}
  \draw [fill] ({\i*0.5*3+1},{\i*sqrt(3)*0.5+sqrt(3)*\j}) circle (1.2mm);
\end{scope}
\begin{scope}
\foreach \i/\j in {4/0,4/1,4/2,4/3,-1/1,-1/2,-1/3,-1/4,-1/0,0/4,1/4,2/4,3/4,0/-1,1/-1,2/-1,3/-1,4/-1}{%
\draw [dashed] ({0.5*3*\i},{0.5*sqrt(3)*\i+sqrt(3)*\j}) -- ({0.5+0.5*3*\i},{(\i+1)*0.5*sqrt(3)+sqrt(3)*\j}) -- ([turn] -60:1) -- ([turn] -60:1) -- ([turn] -60:1) -- ([turn] -60:1) -- cycle; 
}
\end{scope}
\end{tikzpicture}
\caption{A piece of the triangular lattice $\cT$ with the interaction
plaquettes (dotted triangles). The black dots are the vertices of the set 
$\cO\subset \cT$ where the spins are fixed equal to $+1$. The
remaining vertices form an hexagonal lattice
  (thick bonds). Clearly the interaction among the survival spins is a
standard two-body ferromagnetic Ising interaction.}
\label{fig:2}  
\end{figure}
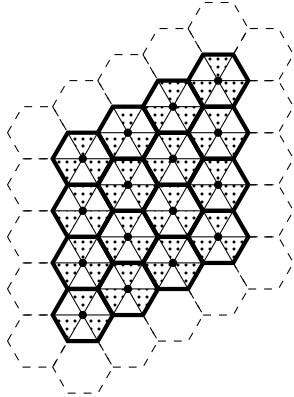
It is
immediate to check that, if we fix  all the spins at
the vertices of $\cO$ equal to $+1$, the remaining spins form a standard Ising model
on the hexagonal grid and the same conclusion valid for the SPM
holds. 
\end{remark}
In the previous section we encountered a first possible critical scale
$\ell_c^{(\text{mix})}$ as the smallest scale at which strong mixing takes
place. 
Here we are going to define two other possible critical
scales and we will prove results about their scaling behaviour in $\b$
as $\b\to \infty$. The first new scale, dubbed
$\ell_c^{(\text{cavity})}$ is the scale at
which the influence of the boundary conditions on bulk
variables (\ie whose support is far from the boundary) is small\footnote{For the experts this is
  essentially the first scale at which \emph{weak mixing} \cite{MO1}
  starts to kick in.}. The
second scale $\ell_c^{(\text{multispin})}$ concerns the smallness of the
infinite volume average of the product of finitely many spins. In what follows the choice of   small numbers $1/10$, $1/5$  is
somewhat arbitrary (one can replace them by $u, 2u$ with $u$ small).  
\begin{definition}
\label{def:length}
Fix two concentric squares $V\subset \L$ of side $\ell$ and $10\ell$
respectively. Given $\t,\t'\in \O$, let $\psi(\ell;\t,\t')$ be the
total variation distance\footnote{Given two probability measures
  $\mu,\nu$ on a finite probability space $\O$ their total variation
  distance is $\max_{A\subset \O}|\mu(A)-\nu(A)|$}  between the marginals on $\O_V$ of the measures
$\mu_\L^\t,\mu_\L^{\t'}$ and let
$\psi(\ell):=\sup_{\t,\t'}\psi(\ell;\t,\t').
$  
Then we set 
\[
\ell_c^{(\text{cavity})}=\min\{\ell:  \ \psi(\ell')\le 1/10 \; \;\forall \ell' \geq \ell \}.
\]
To define the second new scale, let $\bbF_\ell$ be the class of finite
non-empty subsets
$A\subset \bbZ^2$ such that $d(x,y)\ge \ell $ for all $x,y\in A$. Then
we set
\[
  \ell^{(\text{multispin})}_c= \min\{\ell:\ \sup_{A\in \bbF_\ell} |\mu^\b\bigl([\s]_A\bigr)| \le \rosso{1/5}\}.
\]
\end{definition}
\begin{remark}
By the very definition of the variation distance
(cf. \cite{Peres}*{Proposition 4.5}), for $V\subset \L$ as in
Definition \ref{def:length} and for any function $f:\O_\L\mapsto
\bbR$ with $\|f\|_\infty \le 1$ and $S_f\subseteq V$, one has
\[
\max_{\t,\t'}|\mu_\L^{\b,\t}(f)-\mu_\L^{\b,\t'}(f)|\le  \max_{\t,\t'} 2 \psi(\ell;\t,\t')= 2\psi(\ell).
\]  
Therefore \emph{any} observable in $V$ becomes insensitive to the
boundary conditions beyond the scale
$\ell_c^{(\text{cavity})}$. It is a natural question whether there are
special observables for which this phenomenon occurs on a shorter
scale. As discussed in Section \ref{sec:cavity} a very natural
candidate is the spin at the center of the box. In this case a
naive analysis suggests that, under the plus b.c. and for large $\b$, the average spin magnetisation
becomes very small as soon as the side of the box becomes greater than $\bigl(\ell_c^{(\text{cavity})}\bigr)^{1/2}$. In Proposition \ref{prop:magn} we
prove that for the SPM this actually does not happen and that the correct scale is
still $\ell_c^{(\text{cavity})}$. The same question remains open for the TPM.  
\end{remark}

\begin{maintheorem}
\label{thm:main2}
As $\b\to \infty$ the following scaling holds:
\begin{enumerate}[(i)]
\item[(SPM)] 
$\ell_c^{(\text{cavity})}=O(\b \ell_c^{(\text{mix})}),\qquad
\ell_c^{(\text{cavity})}=\O(e^\b),\qquad \ell^{(\text{multispin})}_c\asymp
 e^{\frac{\b}{2}}$. 
\item[(TPM)] 
 $\ell_c^{(\text{cavity})}=O(\b \ell_c^{(\text{mix})}),\qquad \ell_c^{(\text{cavity})}=\O(e^{
   \frac{\ln 2}{ \ln 3} \b} ),\qquad \ell^{(\text{multispin})}_c  \asymp e^{
   \frac{\ln 2}{ \ln 3} \b}$ .  
\end{enumerate}
In particular, for both models $\ell_c^{(\text{cavity})}=O(\b e^\b)$. 
\end{maintheorem}
\begin{remark} \label{rem3.10}
We can use the above result to derive a \emph{lower bound} on
$\ell_c^{(\text{mix})} $. We in fact get $\ell_c^{(\text{mix})}=\O(\b^{-1}e^\b)$ for the
SPM. For the TPM we only get
$\ell_c^{(\text{mix})}=\O(\b^{-1}e^{\frac{\ln 2}{ \ln 3}\b})$. However
one can
prove that the magnetisation of the vertex $x$ at the middle of one of the
sides of
the square of side length $\d e^\b$
is very sensitive to the choice of the boundary conditions on the
other three sides as long as $\d$ is a small constant
independent of $\b$. By using the symmetries defined in Section \ref{sec:6TPM},
one can in fact construct a boundary condition $\t$ such that $\tau$
is identically equal to $+1$ within distance $\d e^\b/2$ from $x$ and
$|m_x^\t - m_x^+|=2m_x^+$, where $m_x^\t,m_x^+$ denote the
magnetisation at $x$ with boundary conditions $\t$ and all plus respectively. In turn
$m_x^+=\O(1)$ as $\b\to \infty$ using Lemma \ref{spiderman}. By
proceeding as in Section \ref{sec:ordering} we easily conclude that 
$\ell_c^{(\text{mix})}=\O(\b^{-1}e^{\b})$. 
We conclude that $\ell_c^{(\text{mix})}$ scales like $e^\beta$, up to polynomial corrections, for both the SPM and TPM. 
\end{remark}
  
\begin{remark} 
As the reader can check, our derivation of Theorem \ref{thm:main1}, as well as our upper bounds on  $\ell_c^{(cavity)}$ and $\ell_c^{(multispin)}$ stated in Theorem \ref{thm:main2}, remain valid when working with generic Hamiltonians of the form  $H(\s) = - \sum _{B \in \cB} J(B) [\s]_B$, where $J(B)$ are arbitrary coupling constants with $\sup\{ |J(B)|\,:\, B \in \cB\} < \infty$. 
\end{remark}

\subsection{Self-similarity of the infinite volume Gibbs measure}
\label{sec:Gibbs}
In this section we establish an exact self--similarity property of the infinite
volume Gibbs measure  $\mu^\beta$.

Let $\varphi (q, k ):= \frac{1}{2}
-\frac{1}{2} (1-2q)^k$ and $q(\b):=
\mu^{\b}([\s]_{B_*}=-1)$. Clearly
$q(\b)=e^{-\b/2}/(e^{-\b/2}+e^{\b/2})$.
\begin{maintheorem}\ 
  \label{thm:main3} 
  \begin{enumerate}[(a)]
  \item For the SPM   the marginal of
$\mu^{\b}$ on $\O_{\ell\bbZ^2}$,  $\ell \in \bbN$,  coincides with $\mu^{\b'}$,
where $\b'=\b'(\b,\ell)$ is such that $q(\beta')= \varphi\bigl( q(\b), \ell^2\bigr)$.
\item For the TPM the marginal of
$\mu^{\b}$ on $\O_{\ell \bbZ^2}$,  for $ \ell = 2^n$ and  $n\in \bbN,$ coincides with $\mu^{\b'}$,
where $\b'$ is such that $q(\beta')=\varphi\bigl( q(\b), \ell^{\a}
\bigr)$ with $\a= \frac{\log 3}{ \log 2}$. 
  \end{enumerate}
  \end{maintheorem}
By explicit computations we have
\begin{equation}\label{eq:main3}
\b'= \log\Bigl( \frac{ 1+ \bigl( 1- 2q(\b) \bigr)^k   }{  1- \bigl( 1- 2q(\b) \bigr)^k   }\Bigr)\quad   \text{ where }  \quad  k =
\begin{cases}
 \ell^2  & \text{ in the SPM}\,,\\
  \ell^\a & \text{ in the TPM}\,.
  \end{cases}
\end{equation}
\subsection{Extensions to trivial factorizable models}
We briefly discuss the extension of Theorems
\ref{thm:main1} and \ref{thm:main2} to an arbitrary \emph{factorizable trivial model} \cites{FS,Slawny3} (FTM in the
sequel) in
$\bbZ^d$. These spin systems have a formal
Hamiltonian like the one introduced in Section \ref{sec:Gibbs
  measure}, namely 
$H(\s)=- \frac 12\sum_{B\in \cB}[\s]_B$, where now a generic plaquette
$B\in \cB$ has the form $B= B_*+x$, with $B_*=
B_*^{(1)}\times\dots\times B_*^{(d)}$ and $B_*^{(i)}\subset
\bbZ$. W.l.o.g. we
can assume that $|B_*^{(i)}|>1$ (otherwise the system breaks into lower dimensional
independent sub-systems) and that $\min\{k\in \bbZ:\ k\in B_*^{(i)}\}=0$. 

Like the SPM also a FTM has a unique Gibbs measure $\mu^\b$ at all
positive temperatures and it satisfies the
Dobrushin-Shlosman uniqueness criterium \cite{FS}. In what follows we
will work with the same length scales $\ell_c^{(\text{mix})},
\ell_c^{(\text{cavity})}$ and $\ell_c^{(\text{multispin})}$ defined
exactly as before and ordered as in \eqref{eq:ordering} (cf. the proof given
in the appendix).   
\begin{maintheorem}
\label{thm:extension}
Choose a FTM. Then as $\b\to \infty$ we have:
\begin{enumerate}[(i)]
\item $\ell_c^{(\text{mix})}=O(\b e^\b)$;
\item $\ell_c^{(\text{cavity})}=O(\b \ell_c^{(\text{mix})})$ and
  $\ell_c^{(\text{cavity})}=\O(e^\b)$;
\item $\ell_c^{(\text{multispin})}=O(e^{\b/d})$. 
\end{enumerate}
 \end{maintheorem}
\subsection{Summary}
In conclusion we observe that the SPM and TPM have (at least) three a priori different
critical length scales. On the first scale covariances in regular
volumes start to decay (exponentially fast) uniformly in the boundary
conditions \emph{and} in the location of the observables. On the
second scale the boundary conditions do not significantly affect the expectation of bulk observables
while on the third scale the product of finitely many spins have small
expectation w.r.t. to the unique infinite volume Gibbs measure. 

The first scale has a scaling roughly $e^{\b}$ with a possible
poly($\b$) pre-factor in both models. The second scale behaves similarly
to the first one in the SPM, whereas for TPM we can only say that it is
in between the first and third scale. Finally the third scale is deeply affected by the
underlying geometry of the models and has a scaling $e^{\b/2}$ for the SPM
and $e^{
   \frac{\ln 2}{ \ln 3} \b}$ for the TPM.   
For both models this third length scale takes the form $e^{\b/d_f}$,
where $d_f $ is the Hausdorff dimension of certain sets which are
naturally associated to the fundamental plaquette $B_*$. 
These sets can be constructed by taking any collection of plaquettes in $\cB$ such that the origin appears in exactly one plaquette and all other vertices of $\bbZ^2$ are include either twice or not at all, and then rescaling space.  
For the SPM this set is just a quadrant in $\bbR^2$, while for the TPM it is given by the Sierpinski gasket. 

One could define a forth length scale as the smallest $\ell$ such that $\b'$ appearing in Theorem \ref{thm:main3} is small, \ie $\b' \leq 1$.
It turns out, by a simple computation from \eqref{eq:main3}, that as expected, this length scales exactly as the multispin length scale.  

\section{Proof of Theorem \ref{thm:main1}}
The proof of Theorem \ref{thm:main1} starts with a simple general
observation. 
Recall the definition of $h_x$ given in Proposition
\ref{thm:fin-size} and let $S_{h_x}$ be its support.
Fix  $\L\subset \bbZ^2$ and $\e\in
(0,1)$. Also fix 
two sites $x,y\notin \L$ such that the $S_x^\L:=S_{h_x}\cap \L$ and
$S_y^\L:=S_{h_y}\cap \L$ are both not empty.
\begin{definition}
\label{def:1}
We say that $\cS\subset \L$ is a $\e$-screening set for $x,y$ if the
following holds:  
\begin{enumerate}[(i)]
\item $\L\setminus \cS$ can be written as the disjoint union of two
sets $V_x,V_y$ such that $S^\L_x\subset V_x$ and $S^\L_y\subset V_y$;
\item there exists no plaquette $B\in \cB(\L)$ such that $B\cap
  V_x\neq \emptyset$ and $B\cap
  V_y\neq \emptyset$;
\item the partition functions $Z_\cS^{\beta,\bullet}$ satisfies the
  bound:
\[
\sup_{\t,\t'} \frac{Z_\cS^{\beta,\t}}{Z_\cS^{\beta,\t'}}\le 1+\e.
\]
\end{enumerate}
\end{definition}
The usefulness of the above definition appears in the next lemma:
\begin{lemma}
\label{lem:1}
In the above setting suppose that there exists an $\e$-screening
set for $x,y$. Then
\[
\sup_\t |\cov_\L^{\b,\t}(h_x,h_y)|\le 4 \|h_x\|_\infty^2\ \e \le 4e^{4\b\|H\|}\ \e.
\]  
\end{lemma}
\begin{proof}
Let $\cS$ be a $\e$-screening set and let $V_x,V_y$ be the
corresponding sets
appearing in Definition \ref{def:1} above. Using the DLR equations we
begin with the simple bound:
\begin{gather*}
  |\cov_\L^{\b,\t}(h_x,h_y)|= |\mu^{\b,\t}_\L\bigl((h_y-\mu_\L^{\b,\t}(h_y))\mu^{\b,\t}_\L(h_x\mid \s_{V_y})\bigr)| \\
\le 
2\|h_y\|_\infty \sup_{\t,\,\xi,\,\xi'}
|\mu^{\b,\t}_{\L}(h_x\mid \s_{V_y}=\xi)-\mu^{\b,\t}_{\L}(h_x\mid \s_{V_y}=\xi')|.
\end{gather*}
Next, given $\xi\in \O_{V_y}$ and using (ii) of Definition \ref{def:1}, we write
\begin{gather*}
  \mu^{\b,\t}_{\L}\left(h_x\mid \s_{V_y}=\xi \right) =
\frac{\sum_{\s\in \O_{V_x}}e^{-\frac{\b}{2} \hat H_{V_x}(\s)} h_x(\s)
  Z_{\cS}^{\b,\t\otimes\s\otimes \xi} }{\sum_{\s'\in \O_{V_x}}e^{-\frac{\b}{2} \hat
    H_{V_x}(\s')}   Z_{\cS}^{\b,\t\otimes\s'\otimes \xi}}, 
\end{gather*}
where
\[
\hat H_{V_x}(\s)= -\sumtwo{B\in \cB(V_x)}{B\cap
  \cS=\emptyset}[\s\otimes \t]_B,
\]
and $\t\otimes\s\otimes \xi$ denotes the configuration whose
restriction to $\L^c, V_x$ and $V_y$ is equal to $\t,\s$ and $\xi$
respectively. Using (iii) of Definition \ref{def:1} we get immediately
that 
\[
(1+\e)^{-1}\le \frac{\mu^{\b,\t}_{\L}\left(h_x\mid
    \s_{V_y}=\xi\right)}{\hat \mu^{\b,\t}_{V_x}(h_x)}\le 1+\e,
\]
where $\hat \mu^{\b,\t}_{V_x}$ is the probability measure on
$\O_{V_x}$ such that  $\hat \mu^{\b,\t}_{V_x}(\s)\propto \exp(-\frac{\b}{2} \hat H_{V_x}(\s))$. Since
$\hat \mu^{\b,\t}_{V_x}(h_x)$ does not depend on $\xi$ we can conclude that
\begin{gather*}
|\mu^{\b,\t}_{\L}(h_x\mid \s_{V_y}=\xi)-\mu^{\b,\t}_{\L}(h_x\mid
\s_{V_y}=\xi')|
\le 2\|h_x\|_\infty\ \e.
\end{gather*}
\end{proof}
For both models $\e$-screening sets will consist of the disjoint union of
suitable translates of the $n$-dilated version of a special set $T_*$. For
the SPM the set $T_*$ will coincide with the fundamental
plaquette $B_*$. For the TPM instead $T_*$ will be the right
triangle with vertices the origin, $\vec e_1$ and $\vec e_1+\vec e_2$. Having that in mind, the next lemma becomes useful.
Before stating it we need to modify a bit the notion of boundary
conditions in order to also cover the case of (partial) free boundary
conditions.

Given a set $\L$ consider a subset $\cP\subset \cB(\L)$ with the
property that $\cP\ni B$ for any plaquette $B\subset \L$. In other
words $\cP$ contains all the plaquettes inside $\L$ and possibly some  of the
``boundary'' plaquettes which intersect $\L$ and $\L^c$. Then we
define 
\begin{equation}
  \label{eq:3}
Z_\L^{\t,\cP}=\sum_{\s\in \O_\L^\t}\exp\bigl(\frac{\b}{2} \sum_{B\in \cP}[\s]_B\bigr).
\end{equation}
\begin{lemma}
\label{lem:2}
Let $\{\L_i\}_{i=1}^k$ be a collection of disjoint subsets of $\bbZ^2$
and let $\L=\cup_{i=1}^k\L_i$. Fix $\e\in (0,1)$ and assume that 
\[
\max_i\max_{\cP}\sup_{\t,\t'}\frac{Z_{\L_i}^{\t,\cP}}{Z_{\L_i}^{\t',\cP}}\le 1+\e.
\]  
Then 
\[
\sup_{\t,\t'} \frac{Z_\L^{\b,\t}}{Z_\L^{\b,\t'}}\le (1+\e)^{k}.
\]
 \end{lemma}
 \begin{proof}
For $i\in [k]$ let $\cP_i=\cB(\L_i)\setminus \{B\in \cB(\L_i): \
\exists j<i \text{ such that } \ B\in \cB(\L_j)\}$. Given $\s^{(i)}\in
\O_{\L_i},\ i\in [k]$, let $\s^{(1)}\otimes\dots\otimes\s^{(k)}\otimes\t$ be
the configuration equal to $\s^{(i)}$ in $\L_i, \ i\in [k],$ and equal to $\t$
outside $\L$. With this notation
\begin{gather*}
  Z_\L^{\b,\t}= \sum_{\s_k\in \O_{\L_k}}\dots \sum_{\s_1\in \O_{\L_1}}
\prod_{i=1}^k\exp\Bigl(\frac{\b}{2}\sum_{B\in \cP_i}[\s^{(1)}\otimes\dots\otimes\s^{(k)}\otimes\t]_B
              \Bigr).
\end{gather*}
Notice that by construction, if $i \ge 2$ then $\sum_{B\in
  \cP_i}[\s^{(1)}\otimes\dots\otimes\s^{(k)}\otimes\t]_B$ does not
depend on $\s^{(1)}\dots \s^{(i-1)}$. 
Thus
\begin{gather*}
  \frac{Z_\L^{\b,\t}}{Z_\L^{\b,\t'}}=\\
= \frac{{\sum_{j=2}^k\sum_{\s^{(j)}\in \O_{\L_j}}}\Bigl[\prod_{i=2}^k\exp\Bigl(\frac{\b}{2}\sum_{B\in \cP_i}[\s^{(i)}\otimes\dots\otimes\s^{(k)}\otimes\t]_B
              \Bigr)\Bigr] Z^{\s^{(2)}\otimes\dots\otimes\s^{(k)}\otimes\t ,\cP_1}_{\L_1}  }{
{\sum_{j=2}^k\sum_{\s^{(j)}\in \O_{\L_j}}\Bigl[\prod_{i=2}^k\exp\Bigl(\frac{\b}{2}\sum_{B\in \cP_i}[\s^{(i)}\otimes\dots\otimes\s^{(k)}\otimes\t']_B
              \Bigr)\Bigr] Z^{\s^{(2)}\otimes\dots\otimes\s^{(k)}\otimes\t'
                ,\cP_1}_{\L_1} }}\\
\le \frac{\max_\t Z^{\t,\cP_1}_{\L_1} }{\min_{\t}Z^{\t,\cP_1}_{\L_1}}\times
\frac{\sum_{j=2}^k\sum_{\s^{(j)}\in \O_{\L_j}}\Bigl[\prod_{i=2}^k\Bigl(\frac{\b}{2}\sum_{B\in \cP_i}[\s^{(i)}\otimes\dots\otimes\s^{(k)}\otimes\t]_B
              \Bigr)\Bigr]}
{\sum_{j=2}^k\sum_{\s^{(j)}\in \O_{\L_j}}\Bigl[\prod_{i=2}^k\exp\Bigl(\frac{\b}{2}\sum_{B\in \cP_i}[\s^{(i)}\otimes\dots\otimes\s^{(k)}\otimes\t']_B
              \Bigr)\Bigr]}\\
\le (1+\e)\times 
\frac{\sum_{j=3}^k\sum_{\s^{(j)}\in \O_{\L_j}}\Bigl[\prod_{i=3}^k\exp\left(\frac{\b}{2}\sum_{B\in \cP_i}[\s^{(i)}\otimes\dots\otimes\s^{(k)}\otimes\t]_B
              \right) \Bigr]Z^{\s^{(3)}\otimes\dots\otimes\s^{(k)}\otimes\t ,\cP_2}_{\L_2}  }{\sum_{j=3}^k\sum_{\s^{(j)}\in \O_{\L_j}}
\Bigl[\prod_{i=3}^k\exp\left(\frac{\b}{2}\sum_{B\in \cP_i}[\s^{(i)}\otimes\dots\otimes\s^{(k)}\otimes\t']_B
              \right) \Bigr]Z^{\s^{(3)}\otimes\dots\otimes\s^{(k)}\otimes\t'
                ,\cP_2}_{\L_2} }.
\end{gather*}
The proof is finished by iteration.
\end{proof}
The next result provides good building blocks of
$\e$-screening sets for large $n$. 
\begin{proposition}
\label{prop:1}
Let $T_*^{(n)}$ be
either the square $[n]^2$ for the SPM or the right triangle with vertices the
origin, $n\vec e_1$ and $n(\vec e_1+\vec e_2)$ for the TPM. Then
\[
\max_\cP\sup_{\t,\t'}\frac{Z^{\t,\cP}_{T_*^{(n)}}}{Z^{\t',\cP}_{T_*^{(n)}}}\le
3 \exp\Bigl(2(n+2) \tanh(\b/2)^{\frac{n+1}{3}}\Bigr) -2,
\]  
when the exponential in the r.h.s is bounded above by $4/3$.
\end{proposition}
The proof of the proposition is postponed to Section 
\ref{sec:proof of prop1}. We now conclude the proof of Theorem \ref{thm:main1} by proving that for both models
$\ell_c^{(\text{mix})}=O(\b e^\b)$ for large $\b$. 

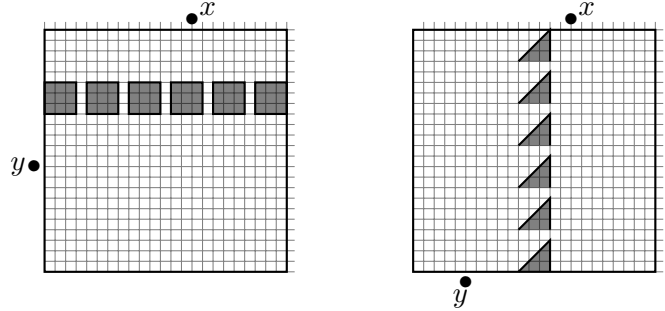
\begin{figure}[ht]
\centering
\begin{tikzpicture}[scale=0.7]
\begin{scope}
\foreach \i in {0,1,2,3,4,5} {%
\draw [thick, fill=gray]  (\i*0.8,3) rectangle
((\i*0.8+0.6,3.6);
}
\draw [semitransparent]  (0,0) grid [step=0.20] (4.75,4.75);
\end{scope}
\draw [thick] (0,0) rectangle (4.6,4.6);
\node at (3.1,5) {$x$};
\node at (2.8,4.8) {$\bullet$};
\node at (-0.5,2) {$y$};
\node at (-0.2,2) {$\bullet$};
\begin{scope}
\foreach \i in {0,1,2,3,4,5} {%
\draw [thick, fill=gray] (9,\i*0.8) -- (9+0.6,\i*0.8+0.6) -- (9.6,\i*0.8);
}
\draw  [semitransparent] (7,0) grid [step=0.20] (7+4.75,4.75);
\end{scope}
\draw [thick] (7,0) rectangle (7+4.6,4.6);
\node at (10.3,5) {$x$};
\node at (10,4.8) {$\bullet$};
\node at (7.9,-0.5) {$y$};
\node at (8,-0.2) {$\bullet$};
\end{tikzpicture}
\caption{Examples of $\e$-screening sets for the SPM (left) and for the TPM
  (right)} 
\label{fig:3}
\end{figure}

$\bullet$ {\it The SPM case.}
Consider the square $Q_\ell$ and $x,y\in Q_\ell^c$ such that
$S_{h_x}\cap Q_\ell\neq \emptyset,\ S_{h_y}\cap Q_\ell\neq \emptyset$
and $d(x,y)\ge \ell/4$. As can be easily checked, taking $\L=Q_\ell$, conditions (i) and (ii) of Definition
\ref{def:1}  are satisfied for suitable sets $V_x,V_y$ when taking $\cS$ equal to either $R_o+z$ or $R_v+z$ for a suitable $z\in \bbZ^2$,  where
$R_o=[\ell]\times [\ell/10]$ and $R_v=[\ell/10]\times[\ell]$,  (w.l.o.g. we assume that $\ell/10\in
\bbN$). Clearly $R_o+z$ (the same for $R_v+z$) can be written as the
disjoint union of ten translates of the square
$Q_{\ell/10}$ (see Figure
\ref{fig:3}). Therefore, if $\ell=c\lfloor\b e^\b\rfloor$ and using
$\tanh(\b/2)\sim \exp(-2e^{-\b})$ for large $\b$, we
obtain from Lemma \ref{lem:2} and Proposition \ref{prop:1}  
\[
     \sup_{\t,\t'}\frac{Z^{\b,\t}_{R_o+z}}{Z^{\b,\t'}_{R_o+z}}\le
     (1+\e(\b))^{10},
\]
with $\e(\b)=O(\b e^{\b(1 - \frac{2c}{30})})$ for $c > 15$. Hence using Lemma
\ref{lem:1} and the same notation of Proposition \ref{thm:fin-size} we get:
\[
\lim_{\b\to \infty} e^{4\b \|H\|}\ell \varphi (\ell) \le
 \lim_{\b\to \infty} C'   \b e^{17\b} \Bigl((1+\e(\b))^{10}-1\Bigr)=0,
\]
for $c$ large enough, where we have used $\|H\| = 2$ for the SPM. 
In particular for the above choice of $\ell$
\eqref{eq:SM} holds for $\b$ large enough.\qed

$\bullet$ {\it The TPM case.}
We proceed exactly as for the SPM. Suppose w.l.o.g. that $R_o+z$ satisfies conditions (i) and (ii) of Definition
\ref{def:1} for the given $x,y$. Then we first divide each of the
ten squares of side $\ell/10$ forming $R_o+z$ into two right triangles
with oblique side along the $\pi/4$ direction and then take as
$\e$-screening set the union of the lowest right triangles (see Figure
\ref{fig:3}). Using again Lemmas \ref{lem:1}, \ref{lem:2} and Proposition
\ref{prop:1} we conclude as in the SPM case.
\subsubsection{Proof of Proposition \ref{prop:1}}
\label{sec:proof of prop1}
 The starting point is a high temperature expansion of the partition
function $Z_\L^{\t,\cP}$ for any finite set $\L$ and 
$\cP\subset \cB(\L)$. Our expansion is similar to but not exactly the
same as the one in \cite{FS} (cf. Eq. (2.9) there). 

Using 
\begin{align*}
  e^{
 (\b/2)  u } = \cosh(\b/2)\left(1+u \tanh(\b/2) \right)\,, \qquad u =\pm 1\,,
\end{align*}
from \eqref{eq:3}
we get
\begin{align}
\label{zebra}
 Z_\L^{\t,\cP}=\cosh(\b/2)^{| \cP| }\Bigl(2^{|\L|} +\sumtwo{\a \subset \cP}{\a\neq \emptyset}\tanh(\b/2)^{|\a|}\sum_{ \s \in \O^\t_\L}\prod_{B\in \a}[\s]_B\Bigr).
\end{align}
\begin{definition}
We say that $\a \subset \cB(\L)$ is a cycle in $\L$ (and write $\a \in
\cK(\L)$) if for each $x \in \L$ the cardinality of $\{ B \in \a \,:\,
x \in B \}$ is even. The empty set  is a cycle.
 \end{definition}
 \begin{remark}
 Notice that since $\cB(\L)$ may contain plaquettes $B$ such that $B\cap
 \L^c\neq \emptyset$, 
 there can be vertices $x\notin \L$
 with odd cardinality of $\{ B \in \a \,:\,
x \in B \}$.
 \end{remark}
If  $\a\subset  \cP$ is not a cycle, then there exists $x \in \L$ such that
the term  $  \prod_{B\in \a}[\s]_B$ can be written as $\s_x
f(\s)$ for some function $f$ with the property that
$x\notin S_f$. Thus  
$\sum_{ \s \in \O_{ \L} }\prod_{B\in \a}[\s]_B=0$. As a consequence we
can write $Z_\L^{\t,\cP}= 2^{|\L|}\cosh(\b/2)^{|\cP|}\cZ_\L^{\t,\cP}\,,\label{creta}
$
 where 
\begin{equation} \label{grecia} 
\cZ_\L^{\t,\cP}:= 1+ 2^{-|\L|} \sum_{\substack{\a \in   \cK(\L)
\\ \a \subset \cP\,,\; \a \neq \emptyset}}\tanh(\b/2)^{|\a|}  \sum_{ \s \in \O^\t_\L }\prod_{B\in \a}[\s]_B\,.
 \end{equation}
Clearly
\[
  |\cZ_\L^{\t,\cP} -1 | \leq \sum_{\substack{\a \in \cK(\L)\\ \a \neq \emptyset}}|\tanh(\b/2)|^{|\a|}\,.
\]
In conclusion if $\sum_{\substack{\a \in \cK(\L)\\ \a \neq
    \emptyset}}|\tanh(\b/2)|^{|\a|}\leq 1/3$ then
\[
\max_{\cP}\sup_{\t,\t'}\frac{\cZ_\L^{\t,\cP}}{\cZ_\L^{\t',\cP}} 
\le 1+3\sum_{\substack{\a \in \cK(\L)\\ \a \neq
    \emptyset}}|\tanh(\b/2)|^{|\a|}.
\]
We now take $\L=T_*^{(n)}$. Then the
 statement of the proposition follows from the following key lemma. 
\begin{lemma}
\label{crimea_ale} 
For both models and for any 
$ t \in (0,1)$ and $n\in \bbN$, 
\begin{equation}\label{gengar}
\sum_{\substack{ \a \in \cK(T_*^{(n)}): \\ \a \not = \emptyset} } t^{|\a|} \leq \exp \{ 2(n+2) t^{\frac{n+1}{3}} \}-1\,.
\end{equation}
\end{lemma}
\begin{proof}[Proof of Lemma \ref{crimea_ale}] \ 

\noindent
$\bullet$ {\it The SPM case.} For any  $0 \leq j \leq n$ call $H_j$ the cycle $\a\in \cK(T_*^{(n)}) $ given by the horizontal stripe of  plaquettes $(k,j)+B_*$, $ 0 \leq k \leq n$.  Define similarly the vertical stripe of plaquettes $V_j=\{ (j,k)+ B_*\,:\, 0 \leq k \leq n\}$.
 \cite{FS}*{Proposition 4.4}  implies that  any cycle $\a\in
 \cK(T_*^{(n)}) $ can be written  as $\a= \a(1)\Delta  \a(2)$, where $\D$ is the symmetric differerence,  $\a(1)$ is a union of cycles of the form $H_j$, $\a(2) $ is a union of cycles of the form $V_j$.
For each $\a\in  \cK(T_*^{(n)})$ we fix once and for all a decomposition $  \a= \a(1)\Delta  \a(2)$ as above, with $\a(1) $ having minimal cardinality. These special decompositions are called   \emph{economic}  in  \cite{FS}, where it is proved that they satisfy the bound $ | \a | \geq \frac{| \a(1) |}{3} + \frac{|\a(2)|}{3}$ (see formula (4.31) there, apart a typo in the sign). Thus we conclude that:
\begin{equation*}
\begin{split}
\sum _{\a \in     \cK(T_*^{(n)})     } t^{|\a|} & \leq 
\sum _{\a \in    \cK(T_*^{(n)})          }  t^{\frac{| \a(1) |}{3}}  t^{\frac{| \a(2) |}{3}} 
 \leq 
   \Big[ \sum _{A \subset \{0,1,\dots, n\}} t^{|A|\frac{n+1}{3}} \Big]^2 =(1+ t^{\frac{n+1}{3}})^{2(n+1)} \,.
\end{split}
\end{equation*}
Using  that $1+x \leq e^x$ we conclude.

\begin{figure}[ht]
  \begin{minipage}{.4\textwidth}
\centering
\newcommand*\rows{9}
\newcommand{\cycle}{(3,0),(3,1),(4,1),(3,2),(5,2),(3,3),(4,3),(5,3),(6,3),(3,4),(7,4)}
\newcommand{\upcycle}{(3,5),(4,5),(7,5),(8,5),(3,6),(5,6),(7,6),(3,7),(4,7),(5,7),(6,7),(7,7),(8,7),(3,8)}

\begin{tikzpicture}[scale=0.43]
[x=1cm, y=1cm]

\foreach \y in {0,...,\rows} {
  \T{$(0,\y)$} \coordinate (a) at (t);
  \T{$(\rows-\y,\rows)$} \coordinate (b) at (t);
  \draw [opacity=0.5] (a) -- (b);
  \T{$(0,\y)$} \coordinate (a) at (t);
  \T{$(\rows,\y)$} \coordinate (b) at (t);
  \draw [opacity=0.5] (a) -- (b);
}
\T{$(0,0)$} \coordinate (a) at (t);
\T{$(0,\rows)$} \coordinate (b) at (t);
\draw [opacity=0.5] (a) -- (b);
\foreach \x in {1,...,\rows} { 
  \T{$(\x,0)$} \coordinate (a) at (t);
  \T{$(\x,\rows)$} \coordinate (b) at (t);
  \draw [opacity=0.5] (a) -- (b);
  \T{$(\x,0)$} \coordinate (a) at (t);
  \T{$(\rows,\rows-\x)$} \coordinate (b) at (t);
  \draw [opacity=0.5] (a) -- (b);
}

\foreach \p [count=\i] in \cycle {
  \coordinate [at=\p, name=P\i];
  \Btri{$(P\i)$};
}
\foreach \p [count=\i] in \upcycle {
  \coordinate [at=\p, name=P\i];
  \Bltri{$(P\i)$};
}

\T{$(3,-0.6)$}
\node at (t)  {$x$};
\draw [thick] (-3,4.3)--(8,4.3);
\node at (7.5,5) {$\cR$};
\node at (3,0){$\bullet$};
\node at (0.5,4.3){$\bullet$};
\node at (1.5,4.3){$\bullet$};
\node at (4.5,4.3){$\bullet$};
\node at (5.5,4.3){$\bullet$};

\end{tikzpicture}
\end{minipage}
  \centering
\ \ 
  \begin{minipage}{.4\textwidth}
    \centering
\newcommand*\rows{9}
\newcommand{\cycle}{(3,0),(3,1),(4,1),(3,2),(5,2),(3,3),(4,3),(5,3),(6,3),(3,4),(7,4)}
\newcommand{\upcycle}{(3,5),(4,5),(7,5),(8,5),(3,6),(5,6),(7,6),(3,7),(4,7),(5,7),(6,7),(7,7),(8,7),(3,8)}

\begin{tikzpicture}[scale=0.4]
[x=1cm, y=1cm]

\foreach \y in {0,...,\rows} {
  \draw [opacity=0.5] (0,\y) -- (\rows-\y,\rows);
  \draw [opacity=0.5] (0,\y) -- (\rows,\y);
}
\draw [opacity=0.5] (0,0) -- (0,\rows);
\foreach \x in {1,...,\rows} { 
  \draw [opacity=0.5] (\x,0) -- (\x,\rows);
  \draw [opacity=0.5] (\x,0) -- (\rows,\rows-\x);
}

\foreach \p [count=\i] in \cycle {
  \coordinate [at=\p, name=P\i];
  \Bsq{$(P\i)$};
}
\foreach \p [count=\i] in \upcycle {
  \coordinate [at=\p, name=P\i];
  \Blsq{$(P\i)$};
}

\T{$(3,-0.6)$}
\node at (t)  {$x$};
\draw [thick] (-0.5,5)--(9.5,5);
\node at (10,5) {$\cR$};
\node at (3,0){$\bullet$};
\node at (3,5){$\bullet$};
\node at (4,5){$\bullet$};
\node at (7,5){$\bullet$};
\node at (8,5){$\bullet$};

\end{tikzpicture}
\end{minipage}

\caption{Left: A piece of the triangular lattice $\cT$ with the line $\cR$
  and the Pascal's triangle $\cP_{x} $ in gray and  light gray.
  The black dots denote the shadow of $x$ on
  $\cR$ and in dark gray are the plaquettes of $\cB_{x,\cR}$. The set consisting of $\{x\}$ and its shadow can be thought of as the sum of all the shadowed plaquettes (dark gray). Right: The same figure on $\bbZ^2$ under the change of basis $\Phi$ (see Remark \ref{rem:TPM}). 
}
\label{fig:5}
\end{figure}
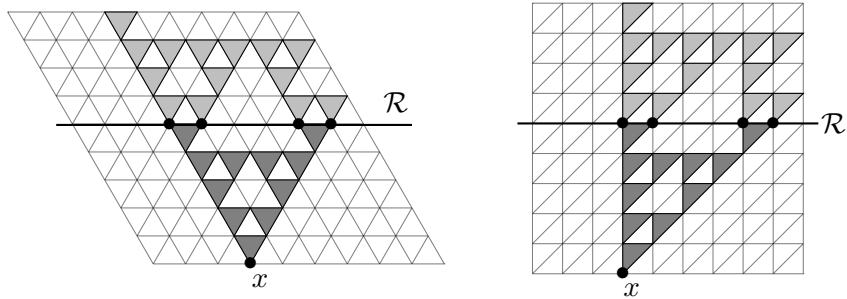
\noindent 
$\bullet$ {\it The TPM case.} Given two sets of plaquettes  $A,B \subset \{ x+B_*\,: \, x \in \bbZ^2\}$, we   define the sum $A+B$ as their symmetric difference $A \D B$. Moreover,  we define the multiplication $k A$ with $k \in \bbF_2=\{0,1\}$ (the field of integers mod 2) as $1 A:=A$ and $0 A:= \emptyset$.   It is then  clear that  the family of subsets of $\cB(T_*^{(n)})$ forms a vector space over $\bbF_2$. 
Let $\cP_x$ be the family of plaquettes belonging to the binary Pascal's triangle rooted at $x$ and developing up-wards on the triangular lattice (see Figure \ref{fig:5}).
For $i=-1, 0, \dots, n$ we denote by $P_i$ the intersection of the plaquettes contained in the Pascal's triangle rooted at $(i,-1)$ with the plaquettes intersecting $T^{(n)}_*$, i.e. (cf. Figure \ref{agosto3})
\[ 
P_i := \cP_{(i,-1)}  \cap \cB(T^{(n)}_*)\,.
\]
  \begin{figure}[tbh]
  \begin{minipage}{.4\textwidth}
    \begin{tikzpicture}[scale=0.4]
      \newcommand{\cycle}{(0,-1),(0,0),(1,0),(2,1),(3,2),(4,3),(5,4),(6,5),(7,6),(8,7),(2,2),(4,4),(6,6),(8,8)}
      \newcommand*\rows{8}
      
      [x=1cm, y=1cm]
      \foreach \y in {-1,...,\rows} {
        \foreach \x in {{\y},...,\rows} { 
          \dtri{$(\x,\y)$};
        }
      }
      \foreach \p [count=\i] in \cycle {
        \coordinate [at=\p, name=P\i];
        \Btri{$(P\i)$};
      }
    
      \T{$(0,0)$} \coordinate (a) at (t);
      \T{$(\rows,\rows)$} \coordinate (b) at (t);
      \T{$(\rows,0)$} \coordinate (c) at (t);
      \draw [ultra thick] (a) -- (b) -- (c) -- cycle;

      \T{$(0,-1)$}
      \node at (t) {$\bullet$};
      \T{$(.75,-2)$}
      \node at (t) {$(0,{-}1)$};
    \end{tikzpicture}
  \end{minipage}
  \centering
  \begin{minipage}{.4\textwidth}
    \centering
    \begin{tikzpicture}[scale=0.4]
      [x=1cm, y=1cm]
      \newcommand*\rows{8}
      \newcommand{\cycle}{(5,-1),(5,0),(5,1),(5,2),(5,3),(5,4),(5,5),(6,0),(7,1),(8,2),(6,2),(7,2),(6,4),(7,5),(6,6),(7,6),(8,6)}
      \foreach \y in {-1,...,\rows} {
        \foreach \x in {{\y},...,\rows} { 
          \dtri{$(\x,\y)$};
        }
      }
      \foreach \p [count=\i] in \cycle {
        \coordinate [at=\p, name=P\i];
        \Btri{$(P\i)$};
      }
      
      \T{$(0,0)$} \coordinate (a) at (t);
      \T{$(\rows,\rows)$} \coordinate (b) at (t);
      \T{$(\rows,0)$} \coordinate (c) at (t);
      \draw [ultra thick] (a) -- (b) -- (c) -- cycle;
      
      \T{$(5,-1)$}
      \node at (t) {$\bullet$};
      \T{$(5.75,-2)$}
      \node at (t) {$(5,{-}1)$};
    \end{tikzpicture}
  \end{minipage}
  \caption{\label{agosto3}
     The region $T^{(8)}_*$ given by all the vertices on and inside the dark lines. The figures show all the plaquettes belonging to $\cB(T^{(8)}_*)$. Left: The plaquettes belonging to $P_0$ are marked in gray. Right: The plaquettes belonging to $P_5$ are marked in gray.
  }
\end{figure}
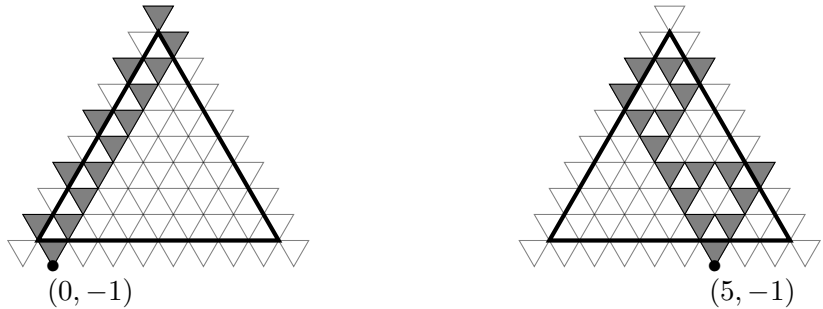
\begin{lemma}\label{ciclomotore} 
The cycle space $\cK(T^{(n)}_*)$ forms a vector space over the field $\bbF_2$ under the symmetric difference. 
Furthermore $P_{-1},P_0,\ldots,P_n$ is a basis.
\end{lemma}
The proof of the lemma is postponed to the appendix.
Using the above lemma we can parametrize all cycles in $\cK(
T^{(n)}_*)$ with $\s \in \{-1,1\}^{[n]_-}$, where $[n]_-=\{-1,0,\ldots,n\}$, via the bijection
\begin{equation}
\label{eq:decomp}
\{-1,1\}^{[n]_-} \ni \s \mapsto \a(\s) \in \cK(T^{(n)}_*), \ \ \textrm{where}\ \  \a(\s) := \sum _{\substack{i\in [n]_-}}\1(\s_i=-1) P_{i} \,.
\end{equation}
Unlike the SPM, we do not have a combinatorial method to efficiently bound from below the cardinality $|\a(\s) |$. 
We will now develop a probabilistic method that turns out to  be effective.
We sample $\s \in \{0,1\}^{[n]_-}$ according to a Bernoulli field of parameter $1/2$, and so
\begin{equation}\label{udine}
\sum_{\substack{ \a \in \cK(T_n^*): \\ \a \not = \emptyset} } t^{|\a|} =
2^{n+2} \bbE \bigl[ t^{|\a(\s) | } \bigr]-1\,.
\end{equation}

In order to control the size of the cycles in \eqref{udine}, we now give a useful characterisation of the plaquettes belonging to a cycle $\a(\s)$ in terms of the product of certain elements of $\s$, which is inspired by a discussion in \cite{Wolfram}.
For this purpose and inspired by \cite{Newman}, it is convenient to introduce the notion of the \emph{shadow of a vertex} in the TMP.
\begin{definition}
\label{def:TPMshadow} 
In the TPM on the triangular lattice $\cT$ let $\cR$ be  an arbitrary horizontal line.
For any vertex $x$ lying below $\cR$ let $\cB_{x,\cR}$ consists of all
plaquettes $B\in \cP_x$ between $x$ and $\cR$ included (recall $\cP_x$ is the set of plaquettes forming an infinite Pascal's
triangle rooted at $x$ rotated by $\pi$).
Then the shadow on $\cR$ of a
vertex $x$ lying on $\cR$ or below it, denoted $S_{x,\cR}$, is $x$
itself in the first case or the set of vertices on  $\cR$  belonging to
an odd number of plaquettes contained in the family $\cB_{x,\cR}$ (cf. Figure
\ref{fig:5}) .
For the TPM on $\bbZ^2$ one simply applies the mapping
$\cT\stackrel {\Phi}{\mapsto}\bbZ^2$ described in Remark
\ref{rem:TPM} to the above geometric construction.
   \end{definition}
 We denote by $\cR_y$ the  horizontal line passing through $(0,y)$, \ie the line at height $y$.
Given $ y \geq -1$, let $S(y) = \{x\in \bbZ\,:\, (x,y) \in
S_{(0,-1),\cR_y}\}$ be the projection on the first coordinate  of the
shadow of the vertex $(0,-1)$  on $\cR_y$ (see Definition
\ref{def:TPMshadow}). We define the set $T^{(n)} $ as 
\[ T^{(n)}:=  \{  z \in \bbZ^2\,:\, z+B_* \in \cB ( T_*^{(n)} ) \} = T_* ^{(n)} \cup \{ (i,-1)\,:\, i \in [n]_-\}\,.
\]
 Given $z \in T^{(n)}$ we define 
\[ \cA(z):= \{ j \in [n]_-\,:\, z+ B_* \in P_j\}\,.\]
Note that by definition of $\a(\s)$ we have
\begin{equation}\label{uffa}
z\in B_* \in \a(\s) \Leftrightarrow [\s]_{\cA (z)}=-1\,. 
\end{equation}
The following claim  characterises the set $\cA(z)$ and describes some of its properties:
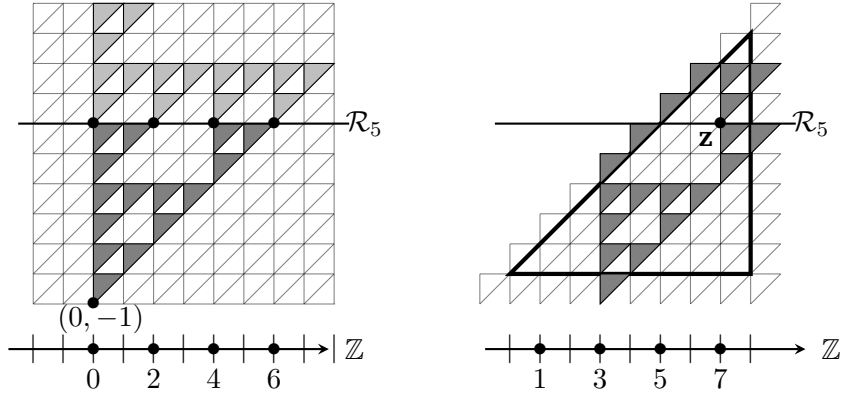
\begin{figure}[tbh]
  \centering
  \begin{minipage}{.4\textwidth}
    \begin{tikzpicture}[scale=0.4,
      axis/.style={thick,->,shorten >=2pt,shorten <=2pt,>=stealth}]
      \centering
      \newcommand*\rows{10}
      \newcommand{\cycle}{(2,0),(2,1),(3,1),(2,2),(4,2),(2,3),(3,3),(4,3),(5,3),(2,4),(6,4),(2,5),(3,5),(6,5),(7,5)}
      \newcommand{\upcycle}{(2,6),(4,6),(6,6),(8,6),(2,7),(3,7),(4,7),(5,7),(6,7),(7,7),(8,7),(9,7),(2,8),(2,9),(3,9)}
      
      [x=1cm, y=1cm]
      \foreach \y in {0,...,\rows} {
        \draw [opacity=0.5] (0,\y) -- (\rows-\y,\rows);
        \draw [opacity=0.5] (0,\y) -- (\rows,\y);
      }
      \draw [opacity=0.5] (0,0) -- (0,\rows);
      \foreach \x in {1,...,\rows} { 
        \draw [opacity=0.5] (\x,0) -- (\x,\rows);
        \draw [opacity=0.5] (\x,0) -- (\rows,\rows-\x);
      }
      
      \foreach \p [count=\i] in \cycle {
        \coordinate [at=\p, name=P\i];
        \Bsq{$(P\i)$};
      }
      \foreach \p [count=\i] in \upcycle {
        \coordinate [at=\p, name=P\i];
        \Blsq{$(P\i)$};
      }
      
      \node at (2.25,-0.5)  {$(0,-1)$};
      \draw [thick] (-0.5,6)--(10.5,6);
      \node at (11,6) {$\cR_5$};
      \node at (2,0){$\bullet$};
      \node at (2,6){$\bullet$};
      \node at (4,6){$\bullet$};
      \node at (6,6){$\bullet$};
      \node at (8,6){$\bullet$};
      
      \draw[axis] (-1,-1.5)  -- (10,-1.5) node(xline)[right]
      {$\bbZ$};
      \foreach \x in {0,...,\rows}{
        \node at (\x,-1.5){$|$};
      }
      \node at (2,-1.5){$\bullet$};
      \node at (2,-2.5)  {$0$};
      \node at (4,-1.5){$\bullet$};
      \node at (4,-2.5)  {$2$};
      \node at (6,-1.5){$\bullet$};
      \node at (6,-2.5)  {$4$};
      \node at (8,-1.5){$\bullet$};
      \node at (8,-2.5)  {$6$};
    \end{tikzpicture}
  \end{minipage}
  \centering
  \begin{minipage}{.4\textwidth}
    \centering
    \begin{tikzpicture}[scale=0.4,
      axis/.style={thick,->,shorten >=2pt,shorten <=2pt,>=stealth}]
      \centering
      \newcommand*\rows{8}
      \newcommand{\cycle}{(3,0),(3,1),(4,1),(3,2),(5,2),(3,3),(4,3),(5,3),(6,3),(3,4),(7,4),(4,5),(7,5),(8,5),(5,6),(7,6),(6,7),(7,7),(8,7)}
      
       [x=1cm, y=1cm]
      \foreach \y in {0,...,\rows} {
        \draw [opacity=0.5] (\y-1,\y+1) -- (\rows+1,\y+1);
      }
      \draw [opacity=0.5] (\rows,\rows+2) -- (\rows+1,\rows+2);
      \draw [opacity=0.5] (-1,0) -- (\rows+1,\rows+2);
      \draw [opacity=0.5] (\rows,0) -- (\rows,\rows +2 );
      \foreach \x in {0,...,\rows} { 
        \draw [opacity=0.5] (\x-1,0) -- (\x-1,\x+1);
        \draw [opacity=0.5] (\x,0) -- (\rows+1,\rows-\x+1);
      }
      \draw [ultra thick] (0,1) -- (\rows,1) -- (\rows,\rows+1) -- cycle;
      
      \foreach \p [count=\i] in \cycle {
        \coordinate [at=\p, name=P\i];
        \Bsq{$(P\i)$};
      }
      \draw [thick] (-0.5,6)--(9.5,6);
      \node at (10,6) {$\cR_5$};
      \node at (7,6){$\bullet$};
      \node at (6.5,5.5){$\textbf{z}$};
      
      \draw[axis] (-1,-1.5)  -- (10,-1.5) node(xline)[right]
      {$\bbZ$};
      \foreach \x in {0,...,\rows}{
        \node at (\x,-1.5){$|$};
      }
      \node at (1,-1.5){$\bullet$};
      \node at (1,-2.5)  {$1$};
      \node at (3,-1.5){$\bullet$};
      \node at (3,-2.5)  {$3$};
      \node at (5,-1.5){$\bullet$};
      \node at (5,-2.5)  {$5$};
      \node at (7,-1.5){$\bullet$};
      \node at (7,-2.5)  {$7$};  
    \end{tikzpicture}
  \end{minipage}
  \caption{\label{agosto6}
 Left: The shadow of the vertex $(0,-1)$ on $\cR_5$ and the construction of the set $S(5)=\{0,2,4,6\}$. Right: a vertex $z=(7,5) \in T_*^{(8)}$, the corresponding set $\cA(z)$ (below), and the cycle $P_3$ in gray. 
  }
\end{figure}
%
%
\begin{claim}\label{harry}
Given $z= (z_1,z_2) \in T^{(n)}$ we have $\cA(z)= z_1 - S(z_2)$. Moreover, we have 
\begin{equation}\label{caio} z_1, z_1-z_2 -1  \in \cA(z) \subset [z_1-z_2-1,z_1]\,.
\end{equation}
\end{claim}
\begin{proof}
Given $ y \geq -1$  it is trivial to check that $S(y) \subset  [0, y+1]$ and $0,y+1 \in S(y)$.
 Writing $z=(z_1,z_2)$, it is clear from the construction that $z + B_* \in P_0$ iff $z_1 \in S(z_2)$.
Fix now a generic $j \in [n]_-$. By translation we have that $z+ B_* \in P_j  $ if and only if  $(z-(j,0)) + B_*\in P_0$, which holds if and only if $z_1 -j \in S(z_2)$, i.e. $j \in z_1 - S(z_2)$.
This proves that 
$ \cA(z)= z_1 - S(z_2)$.

By the initial  observations on $S(y)$ we know that $S(z_2) \subset  [0, z_2+1]$ and $0,z_2+1 \in S(z_2)$. Since $ \cA(z)= z_1 - S(z_2)$,
 we conclude that  that  $\cA(z) \subset [z_1-z_2-1  ,z_1]$ and that $z_1-z_2-1  ,z_1\in \cA(z)$.
\end{proof}
We now introduce special subsets of $T^{(n)} $ denoted by $\G(j)$ with $j=-1,0,1, \dots, n$
(see Figure \ref{agosto7}). $\G(j)$ is defined as 
\begin{align}\label{eq:Gam}
  \G(j) := \{(j,i)\,:\, -1\leq i\leq j\}\cup\{(j+i,i-1)\,:\,1\leq i \leq n-j \}\,.
\end{align}
 \begin{figure}[tbh]
  \centering
  \begin{minipage}{.32\textwidth}
    \centering
    \begin{tikzpicture}[scale=0.4,
      axis/.style={thick,->,shorten >=2pt,shorten <=2pt,>=stealth}]
      \centering
      \newcommand*\rows{8}
      \newcommand{\cycle}{(-1,0),(0,1),(1,2),(2,3),(3,4),(4,5),(5,6),(6,7),(7,8),(8,9)}
      
       [x=1cm, y=1cm]
      \foreach \y in {0,...,\rows} {
        \draw [opacity=0.5] (\y-1,\y+1) -- (\rows+1,\y+1);
      }
      \draw [opacity=0.5] (\rows,\rows+2) -- (\rows+1,\rows+2);
      \draw [opacity=0.5] (-1,0) -- (\rows+1,\rows+2);
      \draw [opacity=0.5] (\rows,0) -- (\rows,\rows +2 );
      \foreach \x in {0,...,\rows} { 
        \draw [opacity=0.5] (\x-1,0) -- (\x-1,\x+1);
        \draw [opacity=0.5] (\x,0) -- (\rows+1,\rows-\x+1);
      }
      \draw [ultra thick] (0,1) -- (\rows,1) -- (\rows,\rows+1) -- cycle;
      
      \foreach \p [count=\i] in \cycle {
        \coordinate [at=\p, name=P\i];
        \Blsq{$(P\i)$};
      }

      \foreach \y in {-1,...,\rows}{
        \node at (\y,\y+1){$\bullet$};
      }
    \end{tikzpicture}
  \end{minipage}
  \centering
  \begin{minipage}{.32\textwidth}
    \centering
    \begin{tikzpicture}[scale=0.4,
      axis/.style={thick,->,shorten >=2pt,shorten <=2pt,>=stealth}]
      \centering
      \newcommand*\rows{8}
      \newcommand{\cycle}{(3,0),(3,1),(4,1),(3,2),(5,2),(3,3),(4,3),(5,3),(6,3),(3,4),(7,4),(4,5),(7,5),(8,5),(5,6),(7,6),(6,7),(7,7),(8,7)}
      
       [x=1cm, y=1cm]
      \foreach \y in {0,...,\rows} {
        \draw [opacity=0.5] (\y-1,\y+1) -- (\rows+1,\y+1);
      }
      \draw [opacity=0.5] (\rows,\rows+2) -- (\rows+1,\rows+2);
      \draw [opacity=0.5] (-1,0) -- (\rows+1,\rows+2);
      \draw [opacity=0.5] (\rows,0) -- (\rows,\rows +2 );
      \foreach \x in {0,...,\rows} { 
        \draw [opacity=0.5] (\x-1,0) -- (\x-1,\x+1);
        \draw [opacity=0.5] (\x,0) -- (\rows+1,\rows-\x+1);
      }
      \draw [ultra thick] (0,1) -- (\rows,1) -- (\rows,\rows+1) -- cycle;
      
      \foreach \p [count=\i] in \cycle {
        \coordinate [at=\p, name=P\i];
        \Blsq{$(P\i)$};
      }

      \foreach \y in {0,...,4}{
        \node at (3,\y){$\bullet$};
      }
     \foreach \x in {4,...,\rows}{
        \node at (\x,\x-3){$\bullet$};
      }
    \end{tikzpicture}
  \end{minipage}
  \centering
  \begin{minipage}{.32\textwidth}
    \centering
    \begin{tikzpicture}[scale=0.4,
      axis/.style={thick,->,shorten >=2pt,shorten <=2pt,>=stealth}]
      \centering
      \newcommand*\rows{8}
      \newcommand{\cycle}{(8,0),(8,1),(8,2),(8,3),(8,4),(8,5),(8,6),(8,7),(8,8),(8,9)}
      
       [x=1cm, y=1cm]
      \foreach \y in {0,...,\rows} {
        \draw [opacity=0.5] (\y-1,\y+1) -- (\rows+1,\y+1);
      }
      \draw [opacity=0.5] (\rows,\rows+2) -- (\rows+1,\rows+2);
      \draw [opacity=0.5] (-1,0) -- (\rows+1,\rows+2);
      \draw [opacity=0.5] (\rows,0) -- (\rows,\rows +2 );
      \foreach \x in {0,...,\rows} { 
        \draw [opacity=0.5] (\x-1,0) -- (\x-1,\x+1);
        \draw [opacity=0.5] (\x,0) -- (\rows+1,\rows-\x+1);
      }
      \draw [ultra thick] (0,1) -- (\rows,1) -- (\rows,\rows+1) -- cycle;
      
      \foreach \p [count=\i] in \cycle {
        \coordinate [at=\p, name=P\i];
        \Blsq{$(P\i)$};
      }

      \foreach \y in {-1,...,\rows}{
        \node at (8,\y+1){$\bullet$};
      }
    \end{tikzpicture}
  \end{minipage}
    \caption{\label{agosto7} The marked vertices show the sets $\G(-1)$ (left), $\G(3)$ (middle) and $\G(8)$(right) for $n=8$. The light gray plaquettes correspond to $P_{-1}$, $P_{3}$ and $P_{8}$ respectively.
}
    \end{figure}
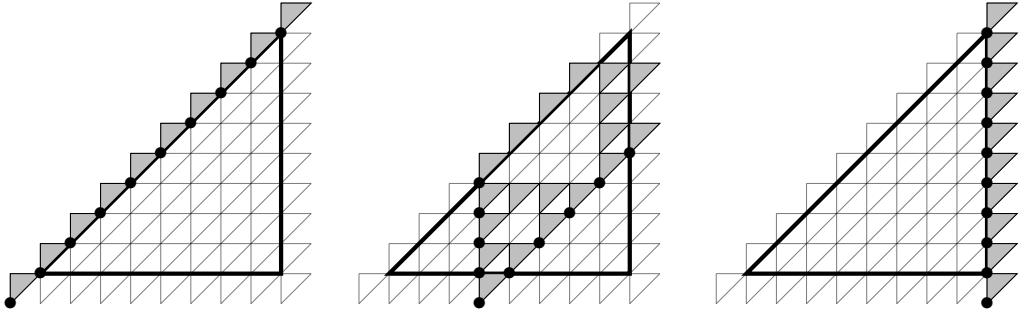
Note that $|\G(j)|=n+2$ for all $ j=-1,0, \dots, n$. 
Moreover, for any $z  \in T^{(n)}$ 
 we have 
\[
\big| \{ j \in [n]_-\,:\, z \in \G(j)\}\big| =
\begin{cases}
1 & \text{ if  $z \in \{(i,-1)\,:\, i\in[n]_-\}$},\\
2 & \text{ otherwise}.
\end{cases}
\]
Now, using these sets and \eqref{uffa}, we can bound the size of each cycle $\alpha(\s) \in \cK(T^{(n)}_*)$ as follows,
\begin{equation}\label{zag}
\begin{split}
| \a(\s) | & = \sum _{z  \in T^{(n)}} \mathds{1} ( z+B_*  \in \a(\s) )  \geq \frac{1}{2} \sum_{j\in [n]_-} \sum_{z  \in \G(j) } \mathds{1} ( z+B_* \in \a (\s) ) 
\\
& =   \frac{1}{2} \sum_{j\in [n]_-} \sum_{z  \in \G(j) } \mathds{1} (  [\s]_{ \cA(z)} =-1 )  \,.
\end{split} 
\end{equation}
It is convenient to write $t= e^{-\e}$ with $\e >0$ (in the applications we have $\e\approx 2e^{-\b}$).
Using \eqref{zag}  and Jensen inequality we can write
\begin{equation}\label{mazinga}
\begin{split}
\bbE \bigl[ t ^{| \a (\s) |} \bigr] & = \bbE \bigl[ e ^{-\e | \a (\s) |} \bigr]
\leq \bbE \Big [ \exp \Big\{ 
   -\frac{\e}{2} \sum_{j\in [n]_-} \sum_{z \in \G(j) } \mathds{1} (  [\s]_{ \cA(z)} =-1 )  \Big\} \Big]\\
   & = \bbE \Big [ \exp \Big\{ 
   -\frac{\e(n+2)}{2} {\rm Av}_j \sum_{z  \in \G(j) } \mathds{1} (  [\s]_{ \cA(z)} =-1 )  \Big \} \Big] \\
   &
   \leq 
   {\rm Av}_j  \bbE \Big [ \exp \Big\{ 
   -\frac{\e(n+2)}{2}  \sum_{z  \in \G(j) } \mathds{1} (  [\s]_{ \cA(z)} =-1 )  \Big\} \Big]   \end{split}
   \end{equation}
To bound the last expectation from above  we need a last and crucial technical fact.
\begin{claim}\label{chiave}
Let $\s\in \{-1,1\}^\bbZ$ be sampled according to the product
Bernoulli$(\frac 12)$ measure. Let $A_1, \dots , A_m$ be finite subsets of $\bbZ$ such that
\begin{equation}\label{pikachu}
A_k \setminus (A_1 \cup A_2\cup \cdots \cup A_{k-1} ) \not = \emptyset \qquad \forall k=1, \dots, m\,.
\end{equation}
Then, for any $c \in \bbR$,
\[ \bbE \Big[ \exp\Big\{ c \sum_{k=1}^m \mathds{1} ( [\s]_{A_k}=-1) \Big\} \Big]= 2^{-m} ( e^c+1)^m \,.
\]
\end{claim}
We will postpone the proof of Claim \ref{chiave} until the end of this section.
We now have all the tools required to conclude the proof of Lemma
\ref{crimea_ale} for the TPM.
In order to apply Claim \ref{chiave} to bound from above the r.h.s. of \eqref{mazinga}, we first show  that we may order the sets $\cA(z)$ for $z\in\G(j)$ such that they satisfy \eqref{pikachu}, \ie each set is nonempty and any set is not included in the union of its  predecessors.
We index the vertices's in  $\G(j)$ as follows (recall \eqref{eq:Gam}),
\begin{align*}
  z^{(i)}=
  \begin{cases}
    (j,-2+i) & \textrm{if } 1 \leq i \leq j+2\,,\\
    (i-2,i-j-3) & \textrm{if } j+3 \leq i \leq n+2\,.
  \end{cases}
\end{align*}
Then the  sets $\cA(z^{(i)})$, $i=1, \dots, n+2$, satisfy condition \eqref{pikachu} by Claim \ref{harry}, in particular  by \eqref{caio}. Indeed, by Claim \ref{harry}, $\cA(z^{(i)})$, $i=1, \dots, n+2$, equals 
\[ j-S(-1)\,, \; j-S(0)\,,\; \dots\,,\, j-S(j) \,, \; j+1-S(0)\,,\; j+2 - S(1) \,,\; \dots\,,\; n-S(n-j-1)\,.
\]
Using \eqref{caio} one gets that the following ordered family of points 
\[  0\,,\; j-1\,,\;j-2\,,\; \dots \,, -1\,,\; j+1\,,\; j+2\,,\; \dots\,, n\] 
has the property the the $i$--th point belongs to $\cA(z^{(i)})$ but it does not belong to  $\cA(z^{(i')})$ for $i' <i$.
 As a consequence,  by applying Claim \ref{chiave}, 
 \begin{equation}\label{zeta}
\bbE \Big [ \exp \Big\{ 
   -\frac{\e(n+2)}{2}  \sum_{z  \in \G(j) } \mathds{1} (  [\s]_{ \cA(z)} =-1 )  \Big\} \Big]    = 
   2^{-n-2} (  e^{- \frac{\e (n+2)}{2} }  +1) ^{n+2}\,.
   \end{equation}
By combining \eqref{udine}, \eqref{mazinga} and \eqref{zeta} we conclude that  (recall that $1+x \leq e^x$)
\begin{equation}
\sum_{\substack{ \a \in \cK(T^{(n)}_*): \\ \a \not = \emptyset} } t^{|\a|} =
2^{n+2} \bbE \bigl[ t^{|\a(\s) | } \bigr]-1 \leq  (  e^{- \frac{\e (n+2)}{2} }  +1) ^{n+2}{-}1 \leq 
\exp \{  (n+2)   e^{- \frac{\e (n+2)}{2} }  \}-1
\end{equation}
as required since $t= e^{- \e}$.
\end{proof}
\begin{proof}[Proof of Claim \ref{chiave}]
We define $G_k := \cup_{r =1}^k A_r$  and define  $\cF_k$ as the $\s$--algebra generated by $\{\s_i\}_{i \in G_k}$. Due to \eqref{pikachu} we can decompose $A_m$ as $A_m=B_m \cup C_m$ with $B_m:= A_m\setminus G_{m-1}\not= \emptyset$ and $C_m := A_m \cap G_{m-1}$, where $B_m$ and $C_m$ are disjoint.
In particular   $[\s]_{A_m}= [\s]_{B_m} [\s]_{C_m}$ and $[\s]_{C_m}$ is $\cF_{m-1}$--measurable.
 Observe that if $\{X_i\}_{i=1}^n$ are i.i.d $\pm 1$
random variables and $q:=\bbP(X_i=-1)$ then $\bbE\left(\prod_{i=1}^n
  X_i\right)=(1-2q)^n$, \ie 
\begin{equation}
\label{margherita} \bbP\bigl(\, \prod_{i=1}^n X_i=1\, \bigr) = \frac{1}{2} + \frac{ (1- 2q)^n}{2}\,.
\end{equation}
Hence, using \eqref{margherita} and $\bbP( \s_i=1)=\bbP( \s_i=-1 )=\frac{1}{2}$, we have
\begin{equation}\label{matteo}
  \bbE\Big[ e^{ c  \mathds{1} ( [\s]_{A_m}=-1) }\,|\, \cF_{m-1}
  \Big ]=
    \bbE\Big[ e^{ c  \mathds{1} ( [\s]_{B_m}=- [\s]_{C_m}) }\,|\, \cF_{m-1}
  \Big ]= \frac{e^c+1}{2}\,,
  \end{equation}
where the last identity follows by integrating over $\s_i $ with $ i \in B_m$.

Note that $[\s]_{A_k}$ is $\cF_{m-1}$--measurable for $k\leq m-1$.  
The by conditioning on $\cF_{m-1}$,  and applying \eqref{matteo},
we get
\begin{equation}\label{nicolo}
\begin{split}
&\bbE \Big[ \exp\Big( c \sum_{k=1}^m \mathds{1} ( [\s]_{A_k}=-1) \Big) \Big]\\
& =
\bbE \Big[ \exp\Big( c \sum_{k=1}^{m-1}  \mathds{1} ( [\s]_{A_k}=-1)\Big)
  \bbE\Big[ \exp \Big(c  \mathds{1} ( [\s]_{A_m}=-1) \Big)\mid \cF_{m-1} 
  \Big ]
 \Big]\\ & = \frac{e^c+1}{2} \bbE \Big[ \exp\Big( c \sum_{k=1}^{m-1} \mathds{1} ( [\s]_{A_k}=-1) \Big) \Big]\,.
\end{split}
\end{equation}
The result now follows by iteration.
\end{proof}

\section{Proof of Theorem \ref{thm:main2}}
Given $\L_1, \L_2, \dots, \L_n\subset \bbZ^2$ we define their sum $\L_1+\dots +\L_n$ as the set of vertices of $\bbZ^2$ belonging to an odd number of $\L_i$'s (\ie for set addition we take the symmetric difference).
Notice that
 \begin{equation}\label{portamivia}
 [\s]_{\L_1+\cdots +\L_n}= \prod _{i=1}^n [\s ]_{\L_i}\,.
 \end{equation}
 We also define $\L,\L'$ to be \emph{equivalent} and write $\L \sim
 \L'$,  if there exist a finite family of plaquettes $\{B_i\}_{i=1}^n$ such that 
$\L= \L'+\sum_{i=1}^n B_i$. When $\L\sim \emptyset$ we say that $\{B_i\}_{i=1}^n$ is a
\emph{plaquettes decomposition} of $\L$ if $\L= B_1+ B_2+ \dots +
B_n$. The decomposition is \emph{minimal} if each $B_i$ appears only
once. It is straightforward to show that the minimal decomposition is unique. 
The importance of the above construction is justified by the following
result proved in \cite{Slawny3}*{Section 4.4}. For any
finite set $A \subset \bbZ^2$ the multispin average $\mu^\beta ( [\s]_A)$ satisfies
\begin{equation}\label{uccellino}
\mu^\beta ( [\s]_A)=\begin{cases}
0 & \text{ if } A \not \sim \emptyset\,,\\
\tanh (\b/2)^n 
& \text{ if } A \sim \emptyset \,,
\end{cases}
\end{equation}
where $n=n(A)$ is the size of the minimal plaquettes decomposition of $A$.
\subsubsection{Asymptotics of
  $\ell_c^{(\text{cavity})}$ as $\b\to \infty$}
\label{sec:cavity}For both models the \emph{upper bound} on $\ell_c^{(\text{cavity})}$ follows
from \eqref{eq:ordering} and Theorem \ref{thm:main1}. For the TPM the \emph{lower
  bound} on  $\ell_c^{(\text{cavity})}$ follows again from
\eqref{eq:ordering} and the lower bound on
$\ell_c^{(\text{multispin})}$ in Theorem \ref{thm:main2}
(cf. below). Thus it remains to prove the lower bound
$\ell_c^{(\text{cavity})}=\O(e^{\b})$ for the SPM. The following simple argument 
forms the basis of our approach. 

Let $A\not\sim \emptyset$ be a finite
set contained inside the square centered at the origin of side
\rosso{$\ell \geq \ell_c^{(\text{cavity})}$}. Then, by definition, for
all boundary conditions $\t$ we must have $|\mu_\L^{\b,\t}([\s]_A)|\le
1/5$, where $\L$ is the square of side  \rosso{ $10 \ell $}
centered at the origin. Using the DLR equations we can write in fact
\begin{gather*}
|\mu_\L^{\b,\t}([\s]_A)|= \big|\int
d\mu^\b(\t')\left(\mu_\L^{\b,\t}([\s]_A)-\mu_\L^{\b,\t'}([\s]_A)\right)\big|\\
\le \int
d\mu^\b(\t')\big
|\mu_\L^{\b,\t}([\s]_A)-\mu_\L^{\b,\t'}([\s]_A)\big|\le
2\int d\mu^\b(\t') \rosso{\psi(\ell ;\t,\t')}\le 1/5,
\end{gather*}
where \rosso{$\psi(\ell;\t,\t')$ }  is the variation
distance introduced in
Definition \ref{def:length}. Thus, in order to bound from below
$\ell_c^{(\text{cavity})}$ by $\ell$, it is enough to prove that for a suitably
chosen boundary
condition $\t$ and $A\subset [-\frac{\ell}{2},\frac{\ell}{2}]^2\cap \bbZ^2$,
we have $|\mu_\L^{\b,\t}([\s]_A)|> 1/5$ if $\L= [-5\ell,5\ell]^2\cap \bbZ^2$. 

Our choice of $\t$ will be the \emph{all plus} boundary conditions. With this
choice it is convenient to define $\cB^+(\L)=\{B\cap \L:\ B\in
 \cB(\L)\}$ and to denote by $B^+$ its generic element. We will write $\cK^+(\L)$ for the family of cycles in
 $\cB^+(\L)$, \ie collections of the $B_i^+$'s such that any point in
 $\L$ belongs to an even number of the $B^+_i$'s. 
\begin{lemma}\label{spiderman} Let $A \subset\L \subset \bbZ^2$ be
 non-empty  finite sets and
suppose that $A= B^+_1+\dots +B^+_n$ with $B^+_i\neq B^+_j$ whenever $ i\neq j.$ Then
\begin{equation*}
 \mu^{\b,+} _\L ([ \s]_A) \geq  \tanh(\b/2)^n \,.
\end{equation*}
\end{lemma}
\begin{proof}
Let  $\a_A:=\{B^+_1,\dots, B^+_n\}$. By applying
\cite{FS}*{eq. (2.12)} together with $ |V_1\triangle V_2|\leq |V_1|+|V_2|$ we get
\begin{equation*}
   \mu^{\b,+} _\L ( \s_A) =\frac{ \sum _{\a \in \cK^+(\L)} \tanh (\b/2)^{|\a\triangle\a_A|} }{ \sum_{\a \in \cK^+(\L)} (\tanh(\b/2)) ^{| \a|} }
   \geq \tanh(\b/2)^{| \a_A| }= \tanh (\b/2) ^{n}.
\end{equation*}
\end{proof}
The above bound is quite crude and one may suspect that for many
choices of $A$ it would be too pessimistic. That is indeed true as we
will see shortly. Nevertheless it is enough to prove the
sought lower bound on $\ell_c^{(\text{cavity})}$. 
    
Choose $\L=[-10\ell, 10\ell]^2\cap \bbZ^2$ and let $A=\{ (0,0),
(0,1)\}$. Clearly $A\not\sim \emptyset$ and $A=B_1^+ +\dots +B^+_{10\ell+1}$ where $B_i^+= B_*+
(i-1,0)$ for $i\in [10\ell]$ and $B^+_{10\ell+1}=
\{(10\ell,0),(10\ell,1)\}$. Using the lemma $\mu^{\b,+} _\L ( \s_A)
\geq \tanh(\b/2)^{10\ell+1}$ \ie $\ell  _c^{\text{(cavity)}}=\O(e^\b)$
by the argument given above.
\qed

The key feature of the above choice of $A$ is that its minimal
plaquettes decomposition requires a \emph{linear} (in $\ell$) number of
plaquettes $B^+$. Another natural choice for $A$ would be the origin
so that $\mu^{\b,+}([\s]_A)$ becomes the magnetisation at the origin
under the plus boundary conditions. In this case the minimal plaquettes
decomposition requires a $\O(\ell^2)$
number of plaquettes $B^+$. Reapplying the above strategy would
however only
produce a lower bound $\O(e^{\b/2})$ on corresponding cavity length. A
natural question is therefore whether the magnetisation at the origin
starts to be (roughly) independent of the boundary conditions on scale
$\approx e^\b$ or on scale $e^{\b/2}$ (or on some scale in between).
The answer is provided in the next result whose technical proof is
deferred to the appendix.
\begin{proposition}
\label{prop:magn}
There exists $c>0$ such that for any $\ell < c e^\b$
\[
\liminf_{\b\to \infty}\mu^{\b,+}_{\L_\ell}(\s_0)>0,
\] 
where $\L_\ell= [-\ell,\ell]^2\cap \bbZ^2$. 
\end{proposition}

\subsubsection{Asymptotics of
  $\ell_c^{(\text{multispin})}$ as $\b\to \infty$}\label{telefonata}
Using \eqref{uccellino} the required bound on $\ell_c^{(\text{multispin})}$
will follow once we are able to estimate the size $n(A)$ of the minimal plaquettes
decomposition of a given finite set $A$. 
For this purpose, we now also define the \emph{shadow of a vertex} for the SPM (recall Definition \ref{def:TPMshadow} for the TPM).
We call positive (negative) \emph{corner} any translation of the set
$\{k\vec e_1:  k\ge 0\}\cup \{k\vec e_2:  k\ge 0\}$ (of the set $\{k\vec e_1:  k\le 0\}\cup \{k\vec e_2:  k\le 0\}$). 
\begin{definition}
\label{def:SPMshadow} For the SPM fix a corner $\cR$ together with $x$ belonging to the quadrant delimited by $\cR$. If $x\in \cR$ then we
define the shadow $S_{x,\cR}$ of $x$ on $\cR$ as
  $x$ itself. Otherwise $S_{x, \cR}$ are the three points on $\cR$ which, together with
  $x$, form the vertices
  of a rectangle. In that case the set of plaquettes contained in the rectangle
  will be denoted by $\cB_{x,\cR}$.

   \end{definition}
The next result gives an algorithmic characterisation of sets $A\sim
\emptyset$ and of their minimal plaquettes decomposition in terms of
the shadows of their elements.
\begin{lemma}
 \label{lem:shadow} 
Let $A\subset \bbZ^2$ be finite and let $\cR$ be either a corner 
(positive or negative) in the SPM or an horizontal line 
in the TPM,
such that every $x\in A$ belongs to the quadrant delimited by $\cR$
in the first case or lies below $\cR$ in the second case.
 Then
 \begin{enumerate}[(i)]
 \item $A \sim \emptyset$ if and only if $\sum _{x \in A} S_{x,\cR}= \emptyset$. 
\item If $A \sim \emptyset$,  
 then the minimal decomposition of $A$ is given by all the plaquettes 
belonging to an odd number of the families $\{\cB_{x,\cR}\}_{x\in A}$
described in Definition \ref{def:TPMshadow} and \ref{def:SPMshadow}.
\end{enumerate}
\end{lemma}
\begin{proof}
(i) Fix a point $x\in A$ and a corner or a line $\cR$ depending on the model such that $x$ casts a shadow on $\cR$.
By definition $\{x\} \sim S_{x,\cR}$ for any $x
\in A$ so that $A= \sum _{x \in A} \{x \}\sim \sum _{x \in A} S_{x,\cR}$.
In order to conclude it is enough to prove that
 $\sum _{x \in A} S_{x,\cR} \sim \emptyset$ iff $\sum _{x \in A} S_{x,\cR}=
 \emptyset$. The ``if '' part is trivial. 
 To prove the opposite
 implication suppose that $\sum _{x \in A} S_{x,\cR}\not = \emptyset$
 while $\sum _{x \in A} S_{x,\cR}\sim \emptyset$. Then there exist $n \geq 1$ distinct plaquettes $B_1, \dots, B_n$ such that  $\sum _{x \in A} S_{x,\cR} =B_1+\cdots +B_n$.
 By a  minimality argument,  one can easily 
 check that the set  $B_1+\cdots +B_n$ cannot belong to  $\cR$, while
 $\sum _{x \in A} S_{x,\cR}$ does, which gives a contradiction.

(ii) Suppose that $A \sim \emptyset$ and notice that $\{x\}= S_{x,\cR} + \sum _{B \in \cB_{x,\cR}} B$. Hence, using $\sum _{x \in A} S_{x,\cR}= \emptyset$ (cf. above), we get
\begin{equation*}
A= \sum_{x\in A}\{x\} = \sum _{x\in A} \bigl( S_{x,\cR} + \sum _{B \in \cB_{x,\cR}} B\bigr)=
 \sum _{x\in A}  \sum _{ B \in \cB_{x,\cR}} B\,.
\end{equation*} 
The r.h.s. in the above expression gives a plaquettes decomposition of
$A$. To get the minimal decomposition it is enough to remove all
possible repetitions.   
\end{proof}
We now have all the
necessary tools to analyse $\ell_c^{(\text{multispin})}$.\\ 

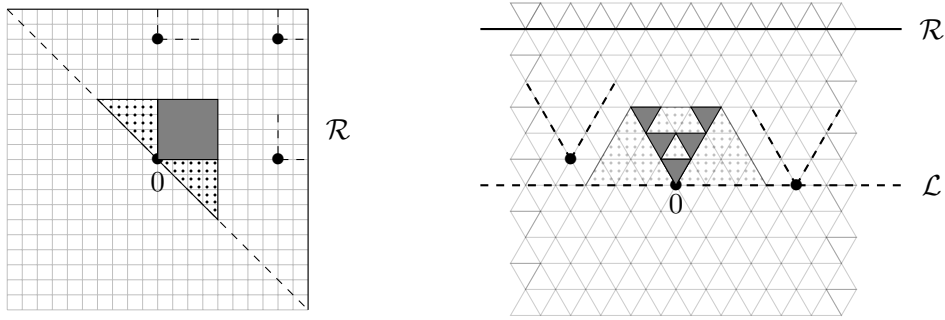
\begin{figure}[ht]
\begin{minipage}{.4\textwidth}
   \centering
\begin{tikzpicture}
\begin{scope}[scale=0.2]
\draw [help lines, semitransparent] (-10,-10) grid (10,10);
\node at (-0,-1.5) {$0$};
\node at (0,0) {$\bullet$};
\node at (8,0) {$\bullet$};
\node at (8,8) {$\bullet$};
\node at (0,8) {$\bullet$};
\node at (12,2) {$\cR$};
\draw [dashed] (-10,10)--(10,-10);
\draw  (-10,10)--(10,10)--(10,-10);
\draw [fill=gray] (0,0)--(4,0)--(4,4)--(0,4)--cycle;
\draw [pattern=dots] (0,0)--(4,0)--(4,-4)--cycle;
\draw [pattern=dots] (0,0)--(0,4)--(-4,4)--cycle;
\draw [dashed] (8,3)--(8,0)--(10,0);
\draw [dashed] (8,10)--(8,8)--(10,8);
\draw [dashed] (0,10)--(0,8)--(3,8);
\end{scope}
\end{tikzpicture}
 \end{minipage}
 \ \hspace{2mm} \hspace{3mm} \
  \centering
\begin{minipage}{.4\textwidth}
   \centering
\newcommand*\rows{5}
\begin{tikzpicture}[scale=0.4]
  \begin{scope}
\foreach \j in {0, 1, ...,\rows} {
\foreach \i in {0,...,10} {
        \draw [thin, opacity=0.2]
        ($(0,0)+\i*(1,0)+2*\j*(0,{0.5*sqrt(3)})$)--($(0.5,{0.5*sqrt(3)})
        +\i*(1,0)+2*\j*(0,{0.5*sqrt(3)})$)--($(1,0)
        +\i*(1,0)+2*\j*(0,{0.5*sqrt(3)})$)--cycle;
\draw [thin,opacity=0.2] ($(0.5,{0.5*sqrt(3)}) +\i*(1,0)+2*\j*(0,{0.5*sqrt(3)})$)--($(1,{sqrt(3)})
+\i*(1,0)+2*\j*(0,{0.5*sqrt(3)})$)--($(1.5,{0.5*sqrt(3)})
+\i*(1,0)+2*\j*(0,{0.5*sqrt(3)})$)-- cycle;
    }
\draw [thin,opacity=0.2] ($(0.5,{0.5*sqrt(3)})+\j*(0,{sqrt(3)})$)--($(0,{sqrt(3)}) +\j*(0,{sqrt(3)})$)--($(1,{sqrt(3)})+\j*(0,{sqrt(3)})$)--cycle;
\draw [thin,opacity=0.2] ($(11,0)+\j*(0,{sqrt(3)})$)--($(10.5,{0.5*sqrt(3)}) +\j*(0,{sqrt(3)})$)--($(11.5,{0.5*sqrt(3)})+\j*(0,{sqrt(3)})$)--cycle;
}
\foreach \col in {0,...,10}{
\draw [thin,opacity=0.2] ($(0.5,{0.5*sqrt(3)})+(0,{10*0.5*sqrt(3)})+\col*(1,0)$)--($(0,{sqrt(3)}) +(0,{10*0.5*sqrt(3)})+\col*(1,0)$)--($(1,{sqrt(3)}) +(0,{10*0.5*sqrt(3)})+\col*(1,0)$)--cycle;
}
\node at ($5.5*(1,{0.5*sqrt(3)})+(0,{-0.6*sqrt(3)})$) {$0$};
\node at (5.5,{5*0.5*sqrt(3)}) {$\bullet$};
\node at (2,{6*0.5*sqrt(3)}) {$\bullet$};
\node at (9.5,{5*0.5*sqrt(3)}) {$\bullet$};
\draw [thick,dashed] (2,{6*0.5*sqrt(3)})--({2+3*0.5},{9*0.5*sqrt(3)});
\draw [thick,dashed] (2,{6*0.5*sqrt(3)})--({2-3*0.5},{9*0.5*sqrt(3)});
\draw [thick,dashed] (9.5,{5*0.5*sqrt(3)})--({9.5+3*0.5},{8*0.5*sqrt(3)});
\draw [thick,dashed] (9.5,{5*0.5*sqrt(3)})--({9.5-3*0.5},{8*0.5*sqrt(3)});
\draw [thick] (-1,{11*0.5*sqrt(3)})--(13,{11*0.5*sqrt(3)});
\node at (14, {11*0.5*sqrt(3)}) {$\cR$};
\draw [dashed, thick] (-1,{5*0.5*sqrt(3)})--(13,{5*0.5*sqrt(3)});
\node at (14,{5*0.5*sqrt(3)}) {$\cL$};
\draw [pattern=dots, semitransparent] (2.5,{5*0.5*sqrt(3)})--(4,{8*0.5*sqrt(3)})--(7, {8*0.5*sqrt(3)})--(8.5,{5*0.5*sqrt(3)});
\foreach \i in {0,1,2}{
 \draw [fill=gray]
 ($(5,0)+\i*(0.5,{0.5*sqrt(3)})+(0.5,{5*0.5*sqrt(3)})$)--($(4.5,{0.5*sqrt(3)})+\i*(0.5,{0.5*sqrt(3)}) +(0.5,{5*0.5*sqrt(3)})$)--%
($(5.5,{0.5*sqrt(3)})+\i*(0.5,{0.5*sqrt(3)}) +(0.5,{5*0.5*sqrt(3)})$)--($(5,0)+\i*(0.5,{0.5*sqrt(3)}) +(0.5,{5*0.5*sqrt(3)})$);
 \draw [fill=gray]
 ($(5,0)+\i*(-0.5,{0.5*sqrt(3)}) +(0.5,{5*0.5*sqrt(3)})$)--($(4.5,{0.5*sqrt(3)})+\i*(-0.5,{0.5*sqrt(3)}) +(0.5,{5*0.5*sqrt(3)})$)--%
($(5.5,{0.5*sqrt(3)}) +(0.5,{5*0.5*sqrt(3)})+\i*(-0.5,{0.5*sqrt(3)})$)--($(5,0) +(0.5,{5*0.5*sqrt(3)})+\i*(-0.5,{0.5*sqrt(3)})$);
 }
 \end{scope}
\end{tikzpicture}
 \end{minipage}
\caption{\emph{Left}: a set $A\sim \emptyset$ in $\bbF_{\ell=8}$ (black dots) above the $-45^o$
   diagonal, the square $W$ (in gray) and the negative corner $\cR$ in the
  construction for the SPM. \emph{Right}: a portion of the set $A\in
  \bbF_{\ell=4}$ (black dots) above the line $\cL$, the
  half-hexagon $\cH$, the Pascal triangle $W$ (in gray) and the line $\cR$ in the
  construction for the TPM. } 
\label{fig:6}
  \label{fig:7}
\end{figure}

\noindent
$\bullet$ {\it The SPM case.} Using \eqref{uccellino} and Lemma \ref{lem:shadow}, if $A\subset
\bbZ^2$ is a square of side length $\ell$, then  $A\sim \emptyset$ and
$\mu^\b ([\s]_A) =\tanh (\b/2)^{\ell^2}$. This implies immediately that 
 $
 \ell^{(\text{multispin})}_c=\O(e^{\b/2})$. 
In order to get an upper bound on $\ell^{(\text{multispin})}_c $, take $A \in
\bbF_\ell$ such that $A\sim \emptyset$ (recall \eqref{uccellino}) and, w.l.o.g., assume that $A$ contains the
origin and is contained in the half-space $\{x\in \bbZ^2:\ x_2\ge
-x_1\}$. Since the minimal distance between the points of $A$ is at
least $\ell$, any  positive
quadrant rooted at $y \in A, \ y\neq 0,$ cannot share plaquettes with 
$W=\{0,\dots, \lfloor \ell/2\rfloor \}^2$ (see Figure  
\ref{fig:6}).
We now take  $\cR$ as  a negative corner such that $A$ and $W$ lies in
the quadrant delimited by $\cR$. Let $B_1,\dots,B_n$ be the minimal
plaquettes decomposition of $A$ w.r.t. $\cR$ as described in Lemma 
\ref{lem:shadow}. By construction all plaquettes 
of $\cB_{0, \cR}$ contained in $W$ cannot belong to any other  family
$\cB_{y, \cR}$, $y \in A\setminus \{0\}$. Hence, $n \geq \lfloor \ell\rfloor^2/4$.  

In conclusion \eqref{uccellino} implies that there exists $c>0$ such that 
\[\mu^\b ([\s]_A)\leq  \tanh (\beta/2) ^{\frac{\lfloor \ell\rfloor^2}{4}} \le
\rosso{  \frac{1}{5}}\,,\qquad \forall \ell \le c\,e^{\b/2}.
\]
\noindent

$\bullet$ {\it The TPM case.} W.l.o.g. and to simplify the notation we consider the TPM on the
triangular lattice $\cT$ with the $\ell_1$-distance replaced by the
graph distance and we define the family $\bbF_\ell$ accordingly. Let
$A$ consists of the vertices $x,y,z$ of a downward pointing equilateral triangle with
side length $2^k$. Using Lemma \ref{lem:shadow} $A\sim
\emptyset$ and its minimal plaquettes decomposition is given by all
the plaquettes lying  in the truncated upward--pointing Pascal's triangle with vertices $x,y,z$.
 The number of  these plaquettes equals $3^k$ while the minimal
 graph distance between points of $A$ is equal to $2^k$. Hence, using
 again \eqref{uccellino}, $\ell^{(\text{multispin})}_c=\O(e^{\b\a})$,
 $\a=\frac{\log 2}{\log 3}$.

To prove the upper bound we proceed similarly to the SPM. Fix
$A\subset \cT$, $A\in \bbF_\ell$ and such that $A \sim \emptyset
$ (recall \eqref{uccellino}). We denote by $k$ the largest integer such that $\ell\ge 2^k$ and, w.l.o.g., we assume that the origin of $\cT$ belongs to $A$ and that
$A$ is contained in the upper half graph delimited by the horizontal
line $\cL$ through the origin. Consider the half-hexagon $\cH$ of side length $2^{k-1}$
centered at the origin and lying above $\cL$ and let $\cR$ be any horizontal line of $\cT$
such that both $A$ and $\cH$ lie below $\cR$. By construction, for any
$x\in A$, $x\neq 0$, the upward Pascal triangle rooted at $x$ cannot
intersect the upward Pascal triangle $W$ rooted at the origin and
contained in $\cH$. Lemma \ref{lem:shadow} implies the
minimal plaquettes decomposition of $A$ must contain all the
plaquettes in $W$. In conclusion the
cardinality of the minimal plaquettes decomposition of $A$ is at least
$3^{k-1}$ and the sought bound $\ell^{(\text{multispin})}_c=O(e^{\b\a})$,
$\a=\frac{\log 2}{\log 3}$, follows as before.
\section{Proof of Theorem \ref{thm:main3}}
 
In order to prove Theorem \ref{thm:main3} it is enough to establish a finite
volume version of the result with free boundary conditions (equations \eqref{eq:3.1} and \eqref{eq:3.1TPM}) 
and use the uniqueness of the thermodynamic limit.\\
Let us start by fixing
some useful notation. Given a finite set
$\L\subset \bbZ^2$ let $\cB^f(\L)=\{B\in \cB(\L): \ B\subset
\L\}$ and define  the Gibbs measure on $\L$ with free boundary conditions as
\[
\mu_\L^{\b,f}(\s)= \frac{e^{\frac{\b}{2} \sum_{B\in
      \cB^f(\L)}[\s]_B}}{ Z^{\beta,f}_\L},\qquad \s \in \O_{\L},
\] 
where $Z_\L^{\beta,f}$ is the normalisation constant.

\subsection{The SPM case}  Given $\ell,N\in \bbN$ let $\L_{\ell,N}= \{0,1,\dots,\ell
N\}^2$ and $\L_{\ell,N}^*= \L_{\ell,N}\cap \ell\bbZ^2$. Our aim is to prove that
\begin{equation}
  \label{eq:3.1}
  \mu^{\b,f}_{\L_{\ell,N}}(\{\sigma: \sigma_{{\L_{\ell,N}^*}}=\eta\})=
  \mu^{\b',f}_{\L_{1,N}}(\eta),\quad \forall\  \eta\in
\O_{\L_{\ell,N}^*},  
\end{equation}
where 
 $\b'=\b'(\b,\ell)$ is as in the statement of Theorem \ref{thm:main3}. 
Above  we have identified  $\eta\in \O_{\L_{\ell,N}^*}$ with a configuration in $\Omega_{\L_{1,N}}$ using the bijection $\Lambda_{1,N}\ni x\mapsto \ell x \in \L_{\ell,N}^*$.

Given $\sigma\in\Omega_{\L_{\ell,N}}$, we denote by $ p_x(\s)$ the $\pm 1$ plaquette
variable $[\s]_{B_x}$ associated to the plaquette
$B_x:=B_*+x$. Notice that the spin configuration $\s\in
  \O_{\L_{\ell,N}}$ is completely determined once we assign its values
  along the South and West boundary of $\L_{\ell,N}$ (\ie the sites
  with at least one coordinate equal to zero)  \emph{and} all the
 plaquette variables associated to $\cB^f(\L_{\ell,N})$. It follows that  
the number of spin configurations compatible with an assignment of
the plaquette variables equals $2^{2N\ell+1}$ and that
\begin{equation}
  \label{eq:3.2}
\mu^{\b,f}_{\L_{\ell,N}}(\s)=\frac{1}{2^{2N\ell+1}}\prod_{x\in
  [0,N\ell-1]^2\cap \bbZ^2}\nu^\b(p_{x}(\s)),
\end{equation}
where $\nu^\b(-1)=q(\beta)$ and $\nu^\b(1)=1-q(\beta)$.

Given $\eta\in\Omega_{\L_{\ell,N}^*}$ we introduce
the  \emph{renormalised} plaquette variables $\hat p_x(\eta)=[\eta]_{\ell B_* +x}$, for $x\in \L_{\ell,N}^*$.
Again $\eta$ is univocally determined by its values on the South and West boundary of $\L_{\ell,N}^*$
and its renormalised plaquette variables.
Notice
also that  it holds \begin{equation}
  \label{eq:3.2bis}\hat p_{x}(\sigma_{\L_{\ell,N}^*})=\prod_{y\in \{0,\dots,\ell -1\}^2+x}p_{y}(\sigma)\qquad\forall \sigma\in\Omega_{\L_{\ell,N}}.\end{equation} 

We can now state the basic lemma concerning the SPM Gibbs measure $\mu_{\L_{\ell,N}}^{\b,f}$:
\begin{lemma}
 \label{lem:key} Fix two  configurations
 $\eta,\eta'$ on $\L_{\ell,N}^*$ such that their
 renormalised plaquette variables coincide. Then  
\[
\mu_{\L_{\ell,N}}^{\b,f} (\{\sigma: \sigma_{{\L_{\ell,N}^*}}=\eta\})=
\mu_{\L_{\ell,N}}^{\b,f} (\{\sigma: \sigma_{{\L_{\ell,N}^*}}=\eta'\}).
\] 
\end{lemma}
\begin{proof}

We introduce a bijection $T$ on $\O_{\L_{\ell,N}}$ as the composition of the bijections $T_x$ with $x$ varying among the sites in the South and West boundary of $\L_{\ell,N}^*$, defined as follows. If $x=(0,x_2)$, then $T_x$ is obtained by flipping all spins on sites with second coordinate $x_2$. If $x=(x_1,0)$ with $x_1>0$, then $T_x$ is obtained by flipping all spins on sites with first coordinate $x_1$.
  To get the thesis it is enough to show that the bijection $T$ satisfies two key properties: \begin{itemize}\item[(i)] 
$\forall \sigma\in \Omega_{\Lambda_{\ell,N}}$ it holds  $\mu_{\L_{\ell,N}}^{\b,f}(\sigma)=\mu_{\L_{\ell,N}}^{\b,f}(T(\sigma))$; 
\item[(ii)]
if $\sigma _{\L_{\ell,N}^*}=\eta$ then $T(\sigma)_{\L_{\ell,N}^*}=\eta'$. \end{itemize}
Property (i) immediately follows noticing that  $p_x(\s)=p_x(T(\s))$ and using \eqref{eq:3.2}. To prove property (ii) we notice, using \eqref{eq:3.2bis}, that 
$T(\s)_{{\L_{\ell,N}^*}}$ has the same renormalised plaquette variables as $\eta$. These in turn coincide with the  renormalised plaquette variables of $\eta'$ by hypothesis. Furthermore 
$T$ has been defined in such a way that, if $\s$
agrees with $ \eta$ on the South-West boundary of $\L_{\ell,N}^*$, then $T(\s)$ agrees with $\eta'$ on the same
set. Therefore property (ii) follows by recalling that the renormalized spin configuration is determined uniquely by its values 
on the South-West boundary of $\L_{\ell,N}^*$, and the renormalised plaquette variables.  \end{proof}
We now claim that for any spin
configuration  $\eta\in\Omega_{\L_{\ell,N}^*}$ it holds
\begin{equation}
  \label{eq:3.4}
\mu^{\b,f}_{\L_{\ell,N}}(\{\sigma: \sigma_{{\L_{\ell,N}^*}}=\eta\})=\frac{1}{2^{2N+1}}\prod_{x\in [0,N\ell-1]^2\cap
  \ell\bbZ^2}\rho^\b(\hat p_{x}(\eta)),
\end{equation}
where 
$\rho^\b(-1)=\varphi(q(\beta),\ell^2))=q(\beta')$
and $\rho^\b(1)=1-\rho^{\beta}(-1)$.
Then \eqref{eq:3.1}  follows by comparing \eqref{eq:3.4} with \eqref{eq:3.2} where in the latter  we take $\ell=1$ and replace
$\beta$ by $\beta'$. 

We are left with proving \eqref{eq:3.4}. It holds
\begin{equation}\label{papavero}
\begin{split}
\mu^{\b,f}_{\L_{\ell,N}} (\{\sigma: \sigma_{{\L_{\ell,N}^*}}=\eta\})&=
\frac{1}{2^{2N+1}}\, \mu^{\b,f}_{\L_{\ell,N}} (\{\sigma: \forall x\,\, \hat p_x(\sigma_{\L_{\ell,N}^*})=\hat p_x(\eta) \})\\& =  
\frac{1}{2^{2N+1}}\prod_{x}\mu^{\b,f}_{\L_{\ell,N}}( \{\sigma:  \prod_{y\in
  \ell B_*+x}p_y(\sigma)=\hat p_x(\eta)\})
\end{split}\end{equation}
where $x$ varies on $[0,N\ell-1]^2\cap
  \ell\bbZ^2$. Indeed, the first equality follows from  Lemma \ref{lem:key} and the fact that the number of configurations $\eta$ 
compatible with  a given choice of 
 renormalised plaquettes variables equals $2^{2N+1}$. The second equality follows from 
  \eqref{eq:3.2bis}. Then 
  \eqref{eq:3.4} is obtained by using equations \eqref{margherita}  and \eqref{papavero}.
 


\subsection{The TPM case}
\label{sec:6TPM}
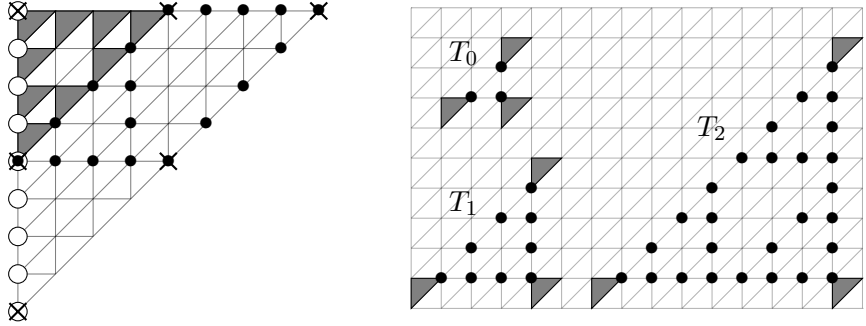
\begin{figure}[thb]
  
   \begin{minipage}{.4\textwidth}
    \centering
\newcommand*\rows{8}
\newcommand{\cycle}{(0,4),(0,5),(1,5),(0,6),(2,6),(0,7),(1,7),(2,7),(3,7)}

\begin{tikzpicture}[scale=0.5]
[x=1cm, y=1cm]
  mycirc/.style={
    circle,
    fill=white,
    draw,
    outer sep=0pt,
    inner sep=1.5pt
  }
\tikzset{cross/.style={
    cross out, 
    draw=black, 
    minimum size=2*(#1-\pgflinewidth), 
    inner sep=0pt, 
    outer sep=0pt},
cross/.default={5pt}}

\foreach \p [count=\i] in \cycle {
  \coordinate [at=\p, name=P\i];
  \Bsq{$(P\i)$};
}


\draw [opacity=0.5] (0,0) -- (0,\rows);
\foreach \x in {1,...,\rows} { 
  \draw [opacity=0.5] (\x,\x) -- (\x,\rows);
}

\foreach \y in {0,...,\rows} {
  \draw [opacity=0.5] (0,\y) -- (\rows-\y,\rows);
  \draw [opacity=0.5] (0,\y) -- (\y,\y);
  \draw [fill=white] (0,\y) circle (7pt); 
}

\draw (0,0) node[cross=4pt,thick] {}; 
\draw (0,4) node[cross=4pt,thick] {}; \draw (4,4) node[cross=4pt,thick] {}; 
\draw (0,8) node[cross=4pt,thick] {}; \draw (4,8) node[cross=4pt,thick] {}; \draw (8,8) node[cross=4pt,thick] {}; 

\node at (0,4){$\bullet$};\node at (1,4){$\bullet$};\node at (1,5){$\bullet$};
\node at (2,4){$\bullet$};\node at (3,4){$\bullet$};\node at (3,5){$\bullet$};
\node at (2,6){$\bullet$};\node at (3,6){$\bullet$};\node at (3,7){$\bullet$};

\node at (4,8){$\bullet$};\node at (5,8){$\bullet$};\node at (6,8){$\bullet$};\node at (7,8){$\bullet$};\node at (8,8){$\bullet$};
\node at (4,4){$\bullet$};\node at (5,5){$\bullet$};\node at (6,6){$\bullet$};\node at (7,7){$\bullet$};

\end{tikzpicture}
\end{minipage}
\centering \ \
\begin{minipage}{.4\textwidth}
    \centering
    
    \newcommand*\rows{10}
    \newcommand*\columns{15}
    
\begin{tikzpicture}[scale=0.4]
[x=1cm, y=1cm]

\foreach \y in {0,...,\rows} {
  \draw [opacity=0.3] (0,\y) -- (\rows-\y,\rows);
  \draw [opacity=0.3] (0,\y) -- (\columns,\y);
}
\draw [opacity=0.3] (0,0) -- (0,\rows);
\foreach \x in {1,...,\columns} { 
  \draw [opacity=0.3] (\x,0) -- (\x,\rows);
  \draw [opacity=0.3] (\x,0) -- ({min(\columns,\x+\rows)},{min(\columns-\x,\rows)});
}

\node at (1.75,8.5) {$T_0$};
\coordinate (P1) at (1,6);
\coordinate (P2) at (3,8);
\coordinate (P3) at (3,6);
\Bsq{$(P1)$};
\Bsq{$(P2)$};
\Bsq{$(P3)$};
\node at (2,7){$\bullet$};
\node at (3,7){$\bullet$};
\node at (3,8){$\bullet$};

\node at (1.75,3.5) {$T_1$};
\coordinate (P1) at (0,0);
\coordinate (P2) at (4,0);
\coordinate (P3) at (4,4);
\Bsq{$(P1)$};
\Bsq{$(P2)$};
\Bsq{$(P3)$};
\node at (1,1){$\bullet$};\node at (2,1){$\bullet$};\node at (2,2){$\bullet$};
\node at (3,1){$\bullet$};\node at (4,1){$\bullet$};\node at (4,2){$\bullet$};
\node at (3,3){$\bullet$};\node at (4,3){$\bullet$};\node at (4,4){$\bullet$};

\node at (10,6) {$T_2$};
\coordinate (P1) at (6,0);
\coordinate (P2) at (14,0);
\coordinate (P3) at (14,8);
\Bsq{$(P1)$};
\Bsq{$(P2)$};
\Bsq{$(P3)$};
\node at (7,1){$\bullet$};\node at (8,1){$\bullet$};\node at (8,2){$\bullet$};
\node at (9,1){$\bullet$};\node at (10,1){$\bullet$};\node at (10,2){$\bullet$};
\node at (9,3){$\bullet$};\node at (10,3){$\bullet$};\node at (10,4){$\bullet$};

\node at (11,1){$\bullet$};\node at (12,1){$\bullet$};\node at (12,2){$\bullet$};
\node at (13,1){$\bullet$};\node at (14,1){$\bullet$};\node at (14,2){$\bullet$};
\node at (13,3){$\bullet$};\node at (14,3){$\bullet$};\node at (14,4){$\bullet$};

\node at (11,5){$\bullet$};\node at (12,5){$\bullet$};\node at (12,6){$\bullet$};
\node at (13,5){$\bullet$};\node at (14,5){$\bullet$};\node at (14,6){$\bullet$};
\node at (13,7){$\bullet$};\node at (14,7){$\bullet$};\node at (14,8){$\bullet$};

\end{tikzpicture}
\end{minipage}

\caption{Left: The region $\cT_{2,1}$. In grey are the plaquettes in $\cP_x^2$ with $x=(0,4)$, crosses correspond to the sites in $\cT_{2,1}^*=\cT_{2,1}\cap 2^2\bbZ^2$, white circles are the sites in $\partial_W \cT_{2,1}$, and the black circles are $T^{(4)}$. Right: Black circles correspond to sites belonging to $T_0$, $T_1$ and $T_2$ (modulo translations). The grey plaquettes are 
those on which the plaquette variable is changed when flipping all spins of $T_i$ (see Claim \ref{keypro}).}
\label{figT}
\end{figure}

We denote by $\mathcal T_{n,N}$ the triangular region in $\mathbb Z^2$ with  vertices the origin, $2^{n+N}\vec e_2$ and $2^{n+N}(\vec e_1+\vec e_2)$ (see Figure \ref{figT} left). Moreover we set $\mathcal T_{n,N}^*:=\mathcal T_{n,N}\cap 2^{n}\bbZ^2$. Our aim is to
prove that
\begin{equation}
  \label{eq:3.1TPM}
  \mu^{\b,f}_{\mathcal T_{n,N}}(\{\sigma: \sigma_{\mathcal T_{n,N}^*}=\eta\})=
  \mu^{\b',f}_{\mathcal T_{1,N}}(\eta),\quad \forall\  \eta\in
\O_{\mathcal T_{n,N}^*}. 
\end{equation}
where we identify $\eta\in \O_{\mathcal T_{n,N}^*}$ with a configuration on $\Omega_{\mathcal T_{1,N}}$ in the natural way.

We define the North and West boundary  of $\mathcal T_{n,N}$ as $\partial_N\mathcal T_{n,N}:=\{(x_1,x_2)\in \mathcal T_{n,N}: x_2=2^{n+N}\}$ and 
 $\partial_W\mathcal T_{n,N}:=\{(x_1,x_2)\in \mathcal T_{n,N}: x_1=0\}$, respectively.
Note that $|\partial_N \cT_{n,N}| = |\partial_W \cT_{n,N}|=\cN(n+N) + 1$, with $\mathcal N(i)=2^i$.

Given $\sigma\in\Omega_{\mathcal T_{n,N}}$, we denote by $p_x(\sigma)$ 
the plaquette variable $ [\s]_{B_*+x}$.
Furthermore, given $\eta\in \O_{\mathcal T_{n,N}^*}$ we  define the \emph{renormalised} plaquette variable $\hat p_x(\eta)=[\eta]_{2^n B_* +x}$, for $x\in \mathcal T_{n,N}^*$.
We observe that the spin configuration $\s \in \O_{\cT_{n,N}}$ is completely determined once we assign its values along the  West boundary \emph{and} the plaquette variables for each plaquette in $\cB^f(\cT_{n,N})$.
It follows that the number of spin configurations compatible with an choice of the plaquette variables is equal to $2^{\cN(n+N)+1}$ and that
\begin{equation}
  \label{eq:probsp}
\mu^{\b,f}_{\cT_{n,N}}(\s)=\frac{1}{2^{\cN(n+N)+1}}\prod_{x\in
  \cT_{n,N}\setminus \partial_N \cT_{n,N}} \nu^\b(p_{x}(\s)),
\end{equation}
where $\nu^\b$ is defined as in the SPM case.
Similarly, $\eta\in \Omega_{\mathcal T_{n,N}^*}$ is univocally determined by its  values  on  $\partial_W \mathcal T_{n,N}^*$ and its renormalised plaquette variables.

Recall that
$\cP_x$ is the family of plaquettes belonging to the binary Pascal's triangle rooted at $x$ and developing up-wards on the triangular lattice (see Figure \ref{fig:5}). 
Then let $\cP_{x}^n$ be the set of plaquettes in $\cP_x$ that are also contained in the triangle with vertices $x,x+2^{n}\vec e_2,x+2^{n}(\vec e_1+\vec e_2)$ (see Figure \ref{figT} left). It is then simple to check that 
\begin{equation}\label{pasca}
\hat p_x(\sigma_{\mathcal T_{n,N}^*})=\prod_{y: y+B_*\in \mathcal \cP_{x}^n}p_{y}(\sigma)\qquad \forall \sigma\in\Omega_{\cT_{n,N}}.
\end{equation}
%

We now define iteratively some geometric sets that we will need to define a bijection on $\Omega_{\cT_{n,N}}$ analogously to what we did in proof of Lemma \ref{lem:key}.
Let $T_0:=\{(0,0), (0,1) ,(1,1)\}$ and 
 $\ell_0:=1$. 
Then  $\forall i\geq 0$ we set
  $$T_{i+1}:=T_{i}\cup \{T_i +(\ell_i+1)\vec e_1\} \cup \{T_{i}+(\ell_i+1)(\vec e_1+\vec e_2)\},\quad\quad \ell_{i+1}:=2\ell_i+1,$$ (which implies $\ell_i=2^{i+1}-1$) as in Figure \ref{figT} (right).
 
The following result can be easily proven inductively:
\begin{claim}\label{keypro}
Fix $i\geq 1$ and $\xi\in\O$. Consider the configuration $\xi'$ obtained  from $\xi$  by flipping all spins in 
$T_i$. Then  the plaquette variables of $\xi$ and $\xi'$ differ at exactly three plaquettes, those with bottom left corner at $(-1,-1)$, $(\ell_i,-1)$ and $(\ell_i,\ell_i)$:
\begin{align*}
  [\xi]_{x+B_*}=
  \begin{cases}
-[\xi]_{x+B_*}  & \textrm{if } x \in\{(-1,-1),(\ell_i,-1),(\ell_i,\ell_i)\},\\
[\xi]_{x+B_*}  & \textrm{otherwise}.
   \end{cases}
\end{align*}

\end{claim}

\begin{lemma}
\label{key}
 Fix two  spin configurations
 $\eta,\eta'$ on $\mathcal T_{n,N}^*$ such that their
 renormalised plaquette variables coincide. Then  
\[
\mu_{\cT_{n,N}}^{\b,f} (\{\sigma:\sigma_{\mathcal T_{n,N}^*}=\eta\})=
\mu_{\cT_{n,N}}^{\b,f} (\{\sigma:\sigma_{\mathcal T_{n,N}^*}=\eta'\}).
\] 
\end{lemma}
\begin{proof}
For $j\in[2^{N+n}]$, we define the bijection $T^{(j)}$ on $\Omega_{\cT_{n,N}}$ as the map that flips all the spins in
$\left( T_{n+N+1}+j\vec e_2 \right)\cap \cT_{n,N}$.
Thanks to Claim \ref{keypro}  the plaquette variables associated to $\cB^f(\cT_{n,N})$ are  left invariant by $T^{(j)}$ and the only spin variable on the West boundary of  ${\cT_{n,N}}$ which is flipped by $T^{(j)}$ is on site $j\vec e_2$.
We define  
 a bijection $T$ on $\O_{\cT_{n,N}}$ as the composition of $T^{(j)}$ for all $j$ such that $\eta_{(0,j)}\neq\eta'_{(0,j)}$.
 Then analogously to the SPM case (cf. proof of Lemma \ref{keypro}) we have (i) $\mu^\b_{\cT_{n,N}}(\s) = \mu^\b_{\cT_{n,N}}(T(\s))$, and (ii) if $\s _{ \mathcal T_{n,N}^*}=\eta$ then $T(\sigma)_{ \mathcal T_{n,N}^*}=\eta'$. Since $T$ is a bijection, the thesis follows 
 from (i) and (ii). 
\end{proof}

It follows from Lemma \ref{key} together with \eqref{pasca}  that for any  $\eta\in\Omega_{\mathcal T_{n,N}^*}$ 
\begin{equation}\label{cognome}
\mu^{\b,f}_{\cT_{n,N}} (\{\sigma:\sigma_{\mathcal T_{n,N}^*}=\eta\})=
\frac{1}{2^{\cN(N)+1}}\prod_{x\in\bar\cT_{n,N}}\mu^{\b,f}_{\cT_{n,N}}(\{\sigma:\prod_{y:y+B_*\in
  \cP^n_x}p_y(\s)=\hat p_x(\eta)\}),
\end{equation}
where $\bar\cT_{n,N}=\mathcal T_{n,N}^* \setminus \partial_N \cT_{n,N}$. 
Then since the number of plaquettes inside the Pascal's triangle of linear size $\ell=2^{n}$ is $\ell^\a=3^{n}$ ($\a= \log 3 / \log 2$), it follows from \eqref{margherita} that
\begin{equation}\label{nome}\mu^{\b,f}_{\cT_{n,N}}(\{\sigma:\prod_{y:y+B_*\in
  \cP^n_x}p_y(\s)=\hat p_x(\eta)\})=\rho^{\b}(\hat p_x(\eta)),\end{equation}
  where
  $\rho^{\beta}(-1)=\varphi(q(\b),\ell^{\alpha})=q(\beta')$ (cf. Theorem \ref{thm:main3}-(b)).
Plugging \eqref{nome} into \eqref{cognome} we get the analogous of \eqref{eq:3.4}.
Due to \eqref{eq:probsp} which is the analogous of \eqref{eq:3.2}, one can derive  
\eqref{eq:3.1TPM} as in the SPM case.

\section{Proof of Theorem \ref{thm:extension}}
\label{sec:extension}
 (i) The proof is identical to the one given for the SPM if one takes
 as $\e$-screening sets large enough $d$-dimensional cubes. For these sets in fact
 the analog of   Lemma \ref{crimea_ale} can be generalised (see also the  arguments in the proof of  
  \cite{FS}*{Proposition 4.5}).

(ii) The upper bound on $\ell_c^{(\text{cavity})}$ follows again from
\eqref{eq:ordering}. To prove the lower bound we follow the pattern of
the proof for the SPM. Let $V\subset \L$ be the two
concentric squares centered at the origin of side $\ell$ and $10\ell$
respectively, and let $A=B_*^{(1)}\times\dots\times B_*^{(d-1)}\times
\{0\}$. For $\ell$ large enough, we have $A \subset V$. It is easy to check that $A\not\sim\emptyset $ so that
$\mu^\b([\s]_A)=0$ (cf.  \cite{Slawny3}*{Section 4.4}).  It is simple to exhibit a subset  $I\subset \{0,1,\dots,10\ell\}$  of cardinality at least  $10\ell/|{\rm diam}(B_*^{(d)})|$ such that
$\{0\}=\sum_{i\in I}(B_*^{(d)}+i)\cap\{0,\dots, 10\ell\}$. Then $A=\sum_{i\in I}(B_*+i\vec
e_d)\cap\L$ and Lemma \ref{spiderman}, which continues to hold in our generality,  implies that 
$\mu_\L^+([\s]_A)\ge \tanh(\b/2)^{|I|}$ \ie
$\ell_c^{(\text{cavity})}=\O(e^\b)$.

(iii) Let  $A \subset \bbZ^d$ be  a finite set such that  $\min\{d(x,y):\ x,y\in A\}\ge \ell$.   W.l.o.g. we assume that $A$ contains the origin. Let $k_i=\max\{j\geq 0\,:\, j \in B_*^{(i)}\}$ for $i\in [d],$ and let $\bbL=\{
\sum_{i=1}^d a_i k_i\vec e_i\,:\ a\in \bbZ^d\}$. Clearly $\bbL$ is
isomorphic to $\bbZ^d$. We set  $A_1=A\cap \bbL$ and 
$A_2=A\setminus A_1$.  We can compute $\mu^\b([\s]_A)$ by first conditioning
on the spins at the vertices of $\bbZ^d\setminus \bbL $ to get 
\[
\mu^\b([\s]_A)=\mu^\b\Bigl([\s]_{A_2}\, \mu^\b\bigl([\s]_{A_1} \mid
\s_{\bbZ^d\setminus \bbL}\bigr)\Bigr).
\]  
Observe now that, given the spins on the vertices of $\bbZ^d\setminus
\bbL$, the law of the remaining spins is again a
FTM with fundamental plaquette $B'_*=\{0,1\}^d$ (upon identifying
$\bbL$ with $\bbZ^d$) and Hamiltonian 
$H'(\eta)=-\frac 12 \sum_{a\in \bbZ^d}J_a[\eta]_{B_*'+a}$, where
$J_a\in \{-1,1\}$ and the sign is determined by the conditioning
spins. 

Using \cite{Szasz} (cf. also \cite{Slawny-rev}*{Formula 4.7}) we get that 
\begin{equation}
\label{eq:ext1}
|\mu^\b\bigl([\s]_{A_1} \mid
\s_{\bbZ^d\setminus \bbL}=\hat \s\bigr)|\le \mu^\b\bigl([\s]_{A_1} \mid
\s_{\bbZ^d\setminus \bbL}\equiv +\bigr),\quad \forall \ \hat \s
\in \{-1,1\}^{\bbZ^d\setminus \bbL}.
\end{equation}
In the r.h.s. of \eqref{eq:ext1} the average is computed w.r.t.  the
\emph{ferromagnetic} trivial factorizable Hamiltonian $H'$ corresponding to
$J_a=+1\ \forall\, a\in \bbZ^d$. In particular it is equal to zero if
$A_1\not\sim \emptyset$ and equal to $\tanh(\b/2)^{n(A_1)}$ otherwise, where the equivalence relation $\sim$ is taken using as
fundamental plaquette the hypercube $B_*'$ and $n(A_1)$ is the
size of the corresponding minimal
plaquette decomposition (see \eqref{uccellino}). By generalising the results of Section \ref{telefonata} for the SPM case, since 
$ A_1\not =\emptyset$ and 
 $\min\{d(x,y):\ x,y\in A_1\}\ge \ell$, one gets that $n(A_1)\ge
c\ell^d$ for some constant $c$ depending on the $k_i$'s. It is
immediate to conclude that 
$\ell_c^{(\text{multispin})}=O(e^{\b/d})$.       
\appendix
\section{}
As anticipated in Section \ref{sec:main results} we collect here the
proof of the ordering of the three length scales
$\ell_c^{(\text{mix})},\ell_c^{(\text{cavity})}$ and
$\ell_c^{(\text{multispin})}$ and of Proposition \ref{prop:magn}. 
\subsection{Proof of (\ref{eq:ordering})}
\label{sec:ordering} We begin by proving that
$\ell_c^{(\text{multispin})}=O(\ell_c^{(\text{cavity})})$. Let
$A$ be a finite set, with $|A| \geq 2$, such that $\min_{x,y\in A}d(x,y)\ge C\ell_c^{(\text{cavity})}$ and fix $x\in A$. 
Let also $\L$ be the square centered at $x$ of side $10\ell_c^{(\text{cavity})}$. For
$C$ large enough no other points of $A$ belong
to $\L$.
Using the DLR equations we write 
\[
\mu^\b([\s]_A)= \int d\mu^\b(\t)\, [\t]_{A\setminus \{x\}}\,\mu_\L^{\b,\t}(\s_x).
\] 
Recall now that $\mu^\b(\s_x)=0$. Thus
\begin{gather*}
|\mu_\L^{\b,\t}(\s_x)|\le \Big|\int
d\mu^\b(\t')\,\left(\mu_\L^{\b,\t}(\s_x)-\mu_\L^{\b,\t'}(\s_x)\right)\Big|
\\
\le
\sup_{\t,\t'}\big|\mu_\L^{\b,\t}(\s_x)-\mu_\L^{\b,\t'}(\s_x)\big|\le \frac{1}{5},
\end{gather*}
by construction and the definition of $\ell_c^{(\text{cavity})}$. The result follows immediately.

Next we prove that
$\ell_c^{(\text{cavity})}=O(\b\,\ell_c^{(\text{mix})})$. Fix
two concentric squares $V\subset \L$ of side $\ell$ and $10\ell$
together with $f:\O_V\mapsto \bbR$ with $\|f\|_\infty\le 1$. The
triangular inequality and \eqref{eq:h.1} (notice that $h_x\ge e^{-2\b
  \|H\|}$) imply that 
\[
\big| \mu_\L^{\b,\t}(f)-\mu_\L^{\b,\t'}(f)\big|\le e^{2\b\|H\|}\sum_{x\in
  \L^c}\sup_{\xi}\big|\cov_\L^{\b,\xi}(h_x,f)\big|. 
\]
Choose now $\ell = C\b\,\ell_c^{(\text{mix})}$. The definition of $\ell_c^{(\text{mix})}$
together with \eqref{eq:SM1} and Proposition \ref{thm:fin-size}
implies that there exists a constant $c$ such that the r.h.s. above is not larger than
$e^{c\b}e^{-\ell/\ell_c^{(\text{mix})}}\ll 1 $ for $C\gg 1$. 
\subsection{Proof of Proposition \ref{prop:magn}}
Let $\L= [-\ell,\ell]^2\cap \bbZ^2$ and recall the definition of the plaquettes family
$\cB^+(\L)$ under plus boundary conditions and of the associated space of cycles
$\cK^+ (\L)$, given right before Lemma \ref{spiderman}. 
In what follows it will useful to think of $\cK^+(\L)$ as  $\bbF_2$--vector space, by  taking the symmetric difference as summation. In particular, if $\cA$ is a collections of
cycles in $\cK^+(\L)$ then a plaquette $B^+$ will belong to
$\sum_{\a\in \cA}\a$ iff
it belongs to an odd number of cycles in $\cA$. 
In $\cK^+(\L)$ we consider the following collection $\cG$ of special cycles
$\{R_i,C_i\}_{i=0}^{2\ell+1}$, called row and column cycles respectively. Order the rows and columns of $\L$ from bottom to top and
left to right. For
$i,j\notin \{0,2\ell+1\}$, the cycle $R_i(C_j)$ consists of all the plaquettes in
$\cB^+(\L)\cap \cB(\L)$ whose
lowermost (leftmost) vertex lies in the $i^{th}$-row ($j^{th}$-column)
of $\L$. For $i=0$ ($i=2\ell+1$) $R_i$ consists of all the plaquettes in
$\cB^+(\L)$ lying on
the first (last) row. Similarly for $C_0,C_\ell$.  

The following are few elementary properties of $\cG$ which we collect
for convenience in a lemma whose proof is omitted.
\begin{lemma}\label{trieste1}\ 
  \begin{enumerate}[(a)]
  \item Let $\cA \subset \cG$ be nonempty and suppose that $\sum _{\a \in \cA} \a =\emptyset$. Then $\cA=\cG$.
\item Let $\cA,\cB \subset \cG$. Suppose that $\sum _{\a\in \cA}\a =\sum_{\a\in \cB}\a$. Then $\cA=\cB$  or $\cB=\cG \setminus \cA$.
\item Given $\a' \in \cK(\L)$ there exists $\cA \subset \cG$ such that
  $\a'=\sum_{\a\in \cA } \a=\sum_{\a\in \cG\setminus \cA } \a$.
  \end{enumerate}
\end{lemma}
\begin{corollary}\label{friuli}
Given $f: \cK^+(\L)\to \bbR$, we have the identity 
\[
\sum _{\a \in \cK^+(\L)} f(\a) = \frac{1}{2} \sum _{ W \subset \cG} f(
\a(W) ),
\] 
where $\a(W):= \sum _{\a \in   W} \a$.
\end{corollary}
We are now ready to write a workable formula for $\mu_\L^+(\s_0)$. Let
$\a_*$ denotes the family of plaquettes in $\cB^+(\L)$ contained in
$[0,\ell]\times [0,-\ell]\cap \bbZ^2$ so that $\sum_{B^+\in \a_*}B^+=\{0\}$.
Using \cite[eq. (2.12)]{FS} and Corollary \ref{friuli} we have
  \begin{equation}\label{viavalvasone}
  \mu^+_\L ( \s_0) = \frac{   \sum _{W \subset \cG}  \tanh(\b/2)^{| \a(W)
      \triangle \a_*|}}{ \sum _{W \subset \cG}  \tanh(\b/2)^{| \a(W) |}  }\equiv
  \frac{N(\b)}{D(\b)}. 
  \end{equation}
\begin{lemma}
If $W\subset \cG$ contains $i$ row cycles and $j$ column cycles then
$|\a(W)|=(i+j)(2\ell +2) -2 ij$. Moreover, if $W\subset \cG$ consists of $u$ rows among $R_0, R_1, \dots, R_\ell$, $v$ rows among $R_{\ell +1}, R_{\ell+2}, \dots, R_{2\ell+1}$, $j$ columns among $C_0, C_1, \dots, C_\ell$ and  $k$ columns  among $C_{\ell +1}, C_{\ell+2}, \dots, C_{2\ell+1}$, then 
\[  |\a(W) \D \a_*|=   jL+vL-2vj-2uj-2vk+2uk\,.\]
\end{lemma}
\begin{proof}
For the first assertion we observe 
that each  $\cB^+(\L)$--plaquette that appears in  $\a(W)$ must be exactly either in a
row cycle or in a column cycle of $W$.  For the second assertion,  we note that  $|\a(W) \D \a_*|$ can be obtained from  $|\a(W)|=(u+v+j+k)(2\ell +2) -2 (u+v)(j+k)$ (as in the first assertion) by subtracting twice the number of $\cB^+(\L)$--plaquettes of $\a(W)$ inside $[0,\ell]\times [0,-\ell]$  (this number is $(u+k)(\ell+1)- 2 uk$) and adding the number  all $\cB^+(\L)$--plaquettes  in $[0,\ell]\times [0,-\ell]$ (this number is $(\ell+1)^2$).
\end{proof}
For notation convenience let $L=2\ell+2$ and let $t=\tanh(\b/2)$. As a consequence of the above lemma we get immediately
\begin{align}\label{ninin1}
D(\b)&= \sum _{i=0}^L \sum _{j=0}^L \binom{L}{i} \binom{L}{j} t^{iL +jL - 2 ij}=
 \sum _{i=0}^L \binom{L}{i}   ( t^i+ t^{L-i})^L\\
N(\b)&= t^{\frac{L^2}{4}}\sum _{u,v,j,k} \binom{L/2}{u} \binom{L/2}{v}
       \binom{L/2}{j} \binom{L/2}{k} t^{ jL+vL-2vj-2uj-2vk+2uk}\\
&= t^{\frac{L^2}{4} } \sum _{u=0} ^{L/2}\sum _{v=0}^{L/2} \binom{L/2}{u} \binom{L/2}{v}( t^{2v}+ t^{L-2v} + t^{2u} + t^{L-2u} ) ^{L/2}.
\end{align}
In order to bound from below the ratio $N(\b)/D(\b)$ it is convenient
to rewrite both expressions in a probabilistic fashion. 
Let $X$ be
a centered $Bin(L,1/2)$ random variable and let $Y,Z$ be i.i.d. centered $Bin(L/2,1/2)$ random variables. Then
\eqref{ninin1} and a little algebra give
\begin{align*}
D(\b)&=2^{2L}t^{L^2/2}\ \bbE\left( \left[ \frac{t^{X}+t^{- X}}{2}\right]^L\right),\\
  N(\b)&=t^{\frac{L^2}{2} }  2^{2L}\bbE\left(\left[\frac{t^{2Y}+t^{-2 Y}+t^{2Z}+t^{-2Z}}{4}\right]^{L/2}\right).
\end{align*}
Trivially $N(\b)\ge 2^{2L}t^{\frac{L^2}{2} }$. Moreover 
$\cosh(x)\le e^{x^2/2}$
implies that 
\[
D(\b)\le 2^{2L}t^{L^2/2}\  \bbE(e^{ (\log t)^2 L X^2/2}).\]
Thus
\[
\frac{N(\b)}{D(\b)}\ge \frac{1}{\bbE(e^{(\log t)^2 L X^2/2})}\,.
\]
Next we bound from above the above expectation value. For any $\l>0$
we write
\begin{gather*}
\bbE(e^{ (\log t)^2 L X^2/2})
\le \l +  \int_{\l}^\infty da \ \bbP(e^{ (\log t)^2 L X^2/2}\ge a).
\end{gather*}
The Azuma-Hoeffding's inequality (cf. e.g. \cite{Mitzen}) implies that $
\bbP(| X|\ge \kappa)\le 2e^{-2\kappa^2/L}$, so that
\[
\bbP(e^{(\log t)^2 L X^2/2}\ge a)\le 
2\exp\left(-4 \,\log a /(\log t)^2 L^2\right)= 2a^{-4/((\log t)^2 L^2)}.
\]
If $\g:=\frac{4}{(\log t)^2 L^2}>1$, then
\begin{equation}
  \label{eq:32}
  \bbE(e^{\log( t)^2 L X^2/2})\le \l + \frac{2}{(\g-1)\l^{\g-1}},
\end{equation}
which, after optimising over the free parameter $\l$, becomes
\[
\bbE(e^{(\log t)^2 L X^2/2})\le 2^{1/\g} +\frac{2}{(\g-1)2^{(\g-1)/\g}}.
\]
Notice that $\g>1$ is equivalent to $L< 2/|\log t| =e^\b(1+o(1))$. 
Recalling that $L=2\ell +2$ we conclude that for any $\d\in (0,1)$ and
any $\ell< (1-\d)e^\b/2$ 
\begin{gather*}
\liminf_{\b\to \infty}  \mu^+_\L ( \s_0) =\liminf_{\b\to \infty}\frac{N(\b)}{D(\b)}>0
\end{gather*}
as required.
\qed

\subsection{Proof of Lemma \ref{ciclomotore}}
To prove the first assertion it is enough to show that if $\a,\g \in \cK( T_*^{(n)})$ then $\a + \g \in \cK( T_*^{(n)})$ . 
To this aim take  $v \in T_*^{(n)}$ and call $\D_1, \D_2, \D_3$ the three triangular plaquettes containing $v$. 
We set $n_i := \mathds{1} ( \D_i \in \a)$ and $m_i:= \mathds{1} ( \D_i \in \g)$. 
Since $\a$ and $\g$ are cycles of $T_n^*$ we have $n_1+n_2+n_3\equiv 0 $ mod 2, and $m_1+m_2+m_3\equiv 0 $ mod 2. 
If we call $k_i := \mathds{1} ( \D_i \in \a+\g)$, by definition of $\a+\g$ we have $k_i = n_i +m_i$ mod 2. We conclude that $k_1+k_2+k_3\equiv 0 $ mod 2, hence $v$ belongs to an even number of plaquettes in $\a+\g$. 
By the arbitrariness of $v \in T_*^{(n)}$ we conclude that $\a+\g$ is a cycle. 

We now prove that $P_{-1},P_0,\ldots,P_n$ is a basis of  $\cK(T^{(n)}_*)$.
We observe that the Pascal's triangle $\cP_0$ (rooted at the origin $0$ of $\bbZ^2$) is a cycle in $\bbZ^2 \setminus \{0\}$, that is every site in $\bbZ^2$ is contained in an even number of plaquettes of $\cP_0$ except for the origin.   
It follows immediately that $P_i$ is a cycle of $T_*^{(n)}$ (see Fig. \ref{fig:5}).
To prove that $P_{-1}, \dots, P_n$ are linearly independent, suppose that $a_{-1} P_{-1}+ \dots+ a_n P_n = \emptyset $ with $a_i \in \bbF_2$. 
By construction we have $(i,-1)+B_* \in P_j$ if and only if $i=j$.
Hence $(i,-1)+B_*$ appears $a_i$ times in the cycle $a_{-1} P_{-1}+ \dots+ a_n P_n = \emptyset $, 
and therefore $a_i=0$ for each $i=-1,0,\ldots,n$.

It remains to show that $P_{-1},P_0,\ldots,P_n$ generates $\cK( T_*^{(n)})$. 
For this we will use the following result:
\begin{claim}\label{miele}
Every non-empty cycle in $\cK( T_*^{(n)}) $ contains a plaquette of the form $(i,-1)+B_*$ for some $i\in\{-1,0,\ldots,n\}$.
\end{claim}
Before proving our claim, we conclude the proof of Lemma \ref{ciclomotore}. 
Fix $\a \in \cK( T_*^{(n)})  $, let $I(\a) = \{i \in \{-1,0,\ldots,n\}\,:\, (i,-1)+B_* \in \a  \}$, and consider the cycle 
 $
\g:= \a + \sum _{ i\in I(\a)    }P_i\, .
$
 By construction $(i,-1)+B_* \not \in \g$ for any $i=-1,0, \dots, n$. 
Therefore, by Claim \ref{miele},  $\g = \emptyset$. 
Equivalently, we have  $\a=\sum _{ i\in I(\a)    }P_i$ as required.  
\begin{proof}[Proof of Claim \ref{miele}]
Suppose, for contradiction, that $\a$ is a non-empty cycle and $(i,-1)+B_* \not \in \a$ for any $i=0, \dots, n$. 
Let $\cR$ be the horizontal line passing through the lowest vertices contained in any plaquette of the cycle $\a$.
By assumption $\cR$ lies on or above the line $\{ k\vec e_1: k \in \bbZ\}$, and therefore $\cR \cap \left( \cup_{B \in \a}B \right)\subset T^{(n)}_*$ and is non empty.
Fix $v\in \cR \cap \left( \cup_{B \in \a}B \right)$, by the definition of a cycle $v$ must belong to an even number of plaquettes in $\a$. 
Since $v$ belongs to exactly one plaquette rooted on $\cR$, and two plaquette rooted below $\cR$, $\a$ must contain at least one plaquette rooted below $\cR$, which contradicts the minimality of $\cR$.  
\end{proof}



\begin{bibdiv}
 \begin{biblist}
 
 
 

\bib{BerthierGarrahanJack}{article}{
author={Berthier,L.},
author={Garrahan, J.P.},
author={Jack,R.},
journal={Physical Review E},
title={Static and dynamic lengthscales in a simple glassy plaquette model},
year={2005},
volume={72}, 
pages={016103-(1-12)}
}


 
\bib{BiroliBerthier}{article}{
author = {Biroli, G.},
author={Berthier, L.},
title = {{Theoretical perspective on the glass transition and amorphous materials}},
journal={Reviews of Modern Physics},
year={2011},
volume={ 83},
pages={587--645}} 


 
\bib{CFM}{article}{
author = {Chleboun, P.},
author={Faggionato, A.},
author={Martinelli, F.},
title = {{Time Scale Separation and Dynamic Heterogeneity in the Low Temperature East Model}},
journal={Commun. Math. Phys.  },
year={2014 },
volume={328 },
pages={ 955--993}}


\bib{DS1}{book}{
author = {Dobrushin, R.},
author={Shlosman, S.},
title = {{Completely Analytical Gibbs Fields}},
booktitle = {Statistical Physics and Dynamical Systems},
year = {1985},
pages = {371--403},
publisher = {Birkh{\"a}user},
address = {Boston, MA}
}

\bib{DS2}{article}{
author = {Dobrushin, R.},
author={Shlosman, S.},
title = {Completely analytical interactions: Constructive description},
journal = {Journal of Statistical Physics},
year = {1987},
volume = {46},
number = {5-6},
pages = {983--1014}
}

 \bib{EspriuPrats}{article}{
author=
{Espriu,D.},
author={Prats,A.},
title={Dynamics of the two-dimensional gonihedric spin model}, 
journal={Physical Review E}, 
volume={70},  
pages={046117-(1-11)},
year={ 2004}}



 
 \bib{EastFMRT}{article}{
author=
{Faggionato,A.},
author={ Martinelli,F.},author={Roberto, C.},author= {Toninelli,C.},
title={ The East model: recent results and new progresses}, journal={Markov Processes and Related Fields}, volume={19}, number={3}, pages={407-452},year={2013}}


\bib{FS}{article}{
author = {Fernandez, R.},
author= {Slawny, J.},
title = {{Inequalities and Many Phase-Transitions in Ferromagnetic Systems}},
journal = {Communications in Mathematical Physics},
year = {1989},
volume = {121},
number = {1},
pages = {91--120}
}


\bib{Garrahanreview}{article}{
author={Garrahan, J.P.},
title={Glassiness through the emergence of effective dynamical constraints in interacting systems}, 
journal={
Journal of Physics: Condensed Matter},
year={2002},
volume={14}, 
pages={1571-1580}
}

\bib{GarrahanJack}{article}{
author={Garrahan, J.P.},
author={Jack,R.},
journal={ The Journal of Chemical Physics},
title={Caging and mosaic length scales in plaquette spin models of glasses},year={2005},volume={123}, pages={164508-(1--14)}}



 \bib{GarrahanSollichToninelli}{book}{
author = {Garrahan, J.P. },
author={Sollich, P.},
author= {Toninelli, C.},
title = {Kinetically Constrained Models},
note={in  ``Dynamical heterogeneities in glasses, colloids, and granular media", Oxford Univ. Press, Eds.: L. Berthier, G. Biroli, J-P Bouchaud, L. Cipelletti and W. van Saarloos (2011)}
}




\bib{MO1}{article}{
author = {Martinelli, F.},
author = {Olivieri, E.},
title = {{Approach to equilibrium of Glauber dynamics in the one-phase region I: the attractive case}},
journal = {Communications in Mathematical Physics},
year = {1994},
volume = {161},
number = {3},
pages = {447--486},
}

\bib{MO2}{article}{
author = {Martinelli, F.},
author = {Olivieri, E.},
title = {{Approach to equilibrium of Glauber dynamics in the one phase
    region. II: the general case}},
journal = {Communications in Mathematical Physics},
year = {1994},
volume = {161},
number = {3},
pages = {487--514}
}


\bib{MO3}{article}{
author = {Martinelli, F.},
author = {Olivieri, E.},
title = {{Finite Volume Mixing Conditions for Lattice Spin Systems and Exponential Approach to Equilibrium of Glauber Dynamics}},
booktitle = {Cellular Automata and Cooperative Systems},
year = {1993},
pages = {473--490},
publisher = {Springer Netherlands},
address = {Dordrecht}
}

\bib{MOS}{article}{
author = {Martinelli, F.},
author = {Olivieri, E.},
author={Schonmann,R.},
title = {{For 2-D lattice spin systems weak mixing implies strong mixing}},
journal = {Communications in Mathematical Physics},
year = {1994},
volume = {165},
number = {1},
pages = {33--47}
}


 
 
 

 
 
 
 
 
 
 
 


 
\bib{M}{book}{
author = {Martinelli, F.},
editor={Minlos,R.},
editor={Shlosman, S.},
editor={Suhov, Yu. M.},
title = {An elementary approach to finite size conditions for the exponential decay of covariances in lattice spin models},
series = {American Mathematical Society Translations: Series 2},
 publisher = {American Mathematical Society},
year = {2000},
volume = {198},
url={},
}


\bib{Mitzen}{book}{
author={Mitzenmacher, M.},
author= {Upfal,E.},
title={Probability and computing},
publisher={Cambridge University Press},
date={2005},
ISBN={0-521-83540-2}
}



\bib{Mueller}{article}{
author=
{Mueller,M.},
author={Johnston, D. A.},
author={Janke,W.},
title={Exact solutions to plaquette Ising models with free and periodic boundaries }, 
journal={Nuclear Physics B},
volume={914},  
pages={388--404},
year={2017}
}



\bib{Newman}{article}{
author = {Newman, M. E. J.},
author={ Moore, C.},
title = {{Glassy dynamics and aging in an exactly solvable spin model}},
journal = {Physical Review E},
year = {1999},
volume = {60},
number = {5},
pages = {5068--5072},
}



\bib{Peres}{book}{
      author={Levin, David~A.},
author={Peres, Y.},
author={Wilmer, E. L.},
       title={Markov chains and mixing times},
   publisher={American Mathematical Society},
        date={2009},
        ISBN={978-0-8218-4739-8},
        note={With a chapter by James G. Propp and David B. Wilson},
}

\bib{Sh}{article}{
author = {Shlosman, S.B.},
title={Uniqueness and Half Space Nonuniqueness of Gibbs States in
  Czech Models},
journal = {Theoretical and Mathematical Physics},
year ={1986},
volume ={66}, 
pages = {284--293},
} 

\bib{Slawny-rev}{book}{
editor={Domb, C.},
editor={Lebowitz, J.L.},
author = {Slawny, J.},
title = {{Low Temperature properites of Classical lattice
    Systems: Phase Transitions and Phase Diagrams }},
series={Phase transitions and critical phenomena},
volume={11},
year = {1987},
pages = {128--202},
}

\bib{Slawny3}{article}{
author = {Holsztynski, W.},
author={Slawny, J.},
title = {{Phase transitions in ferromagnetic spin systems at low temperatures}},
journal = {Communications in Mathematical Physics},
year = {1979},
volume = {66},
number = {2},
pages = {147--166}
}
\bib{Szasz}{article}{,
author = {Sz\`asz, D.},
title = {{Correlation inequalities for non-purely-ferromagnetic systems}},
journal = {Journal of Statistical Physics},
year = {1978},
volume = {19},
number = {5},
pages = {453--459}
}
\bib{Wolfram}{article}{
author = {Wolfram, S.},
title = {Statistical mechanics in cellular automata},
journal = {Reviews of Modern Physics},
year = {1983},
volume = {55},
pages = {601-644}
}





 \end{biblist}
 \end{bibdiv}
\end{document}